\documentclass[preprint]{elsarticle}
\usepackage{lineno,hyperref}
\modulolinenumbers[5]

\usepackage{tabulary}
\usepackage{booktabs}
\usepackage{multirow}

\begin{document}

\begin{frontmatter}

\title{The Plastic Scintillator Detector at DAMPE}
%%\tnotetext[mytitlenote]{Fully documented templates are available in the elsarticle package on %%\href{http://www.ctan.org/tex-archive/macros/latex/contrib/elsarticle}{CTAN}.}


\author[imp]{Yuhong Yu}

\author[imp]{Zhiyu Sun\corref{corresponding_author}}
\cortext[corresponding_author]{Corresponding author}
\ead{sunzhy@impcas.ac.cn}

\author[imp]{Hong Su}
\author[imp]{Yaqing Yang}

\author[imp]{Jie Liu\corref{corresponding_author}}
\ead{j.liu@impcas.ac.cn}

\author[imp]{Jie Kong}
\author[imp]{Guoqing Xiao}
\author[imp]{Xinwen Ma}
\author[imp]{Yong Zhou}
\author[imp]{Hongyun Zhao}
\author[imp]{Dan Mo}

\author[imp]{Yongjie Zhang}
\author[imp,ucas]{Peng Yang}

\author[imp]{Junling Chen}
\author[imp]{Haibo Yang}
\author[imp]{Fang Fang}
\author[imp]{Shengxia Zhang}

\author[imp]{HuiJun Yao}
\author[imp]{Jinglai Duan}
\author[imp]{Xiaoyang Niu}
\author[imp]{Zhengguo Hu}
\author[imp,ucas]{Zhaomin Wang}
\author[imp]{Xiaohui Wang}
\author[imp]{Jingzhe Zhang}
\author[imp]{Wenqiang Liu}

\address[imp]{Institute of Modern Physics, Chinese Academy of Sciences,  509 Nanchang Road,  Lanzhou,  730000,  P.R.China}

\address[ucas]{Graduate University of the Chinese Academy of Sciences,  19A Yuquan Road,  Beijing,  100049,  P.R.China}

\begin{abstract}

The DArk Matter Particle Explorer (DAMPE) is a general purposed satellite-borne high energy $\gamma-$ray and cosmic ray detector, and among the scientific objectives of DAMPE are the searches for the origin of cosmic rays and an understanding of Dark Matter particles. As one of the four detectors in DAMPE, the Plastic Scintillator Detector (PSD) plays an important role in the particle charge measurement and the photons/electrons separation.
The PSD has 82 modules, each consists of a long organic plastic scintillator bar and two PMTs at both ends for readout, in two layers and covers an overall active area larger than 82 cm $\times$ 82 cm. It can identify the charge states for relativistic ions from H to Fe, and the detector efficiency for Z=1 particles can reach 0.9999. The PSD has been successfully launched with DAMPE on Dec. 17, 2015. In this paper, the design, the assembly, the qualification tests of the PSD and some of the performance measured on the ground have been described in detail.

\end{abstract}


\end{frontmatter}

%%\linenumbers

%%%%%%%%%%%%%%%%  Introduction  %%%%%%%%%%%%%%%%%%%%%%%
\section{Introduction}

Although the existence of Dark Matter (DM) has been proved by many astronomical observations~\cite{Rubin,Clowe}, we still know very little about the DM and have many questions, such as its composition. Because it cannot be explained within the framework of the standard model in particle physics, searching for the dark matter particle is a hotspot in today's physics studies, which may probably lead to a new revolution in existing physics theories.

There is a long list that scientists try to find the answer, either in space or deeper under the ground, and no clear signal has been found till now~\cite{Klasen}.

The DArk Matter Particle Explore (DAMPE) is a new one in the list, and it's a satellite-borne high energy particle detector supported by the Strategic Pioneer Program on Space Science of the Chinese Academy of Sciences (CAS)~\cite{chang}.

As a high energy particle detector, DAMPE is able to measure electron, photon, proton, helium and other heavy ions in wide energy ranges with good energy resolution and large acceptance. The primary goal of DAMPE is to probe the nature of dark matter. It will identify possible DM signatures by measuring electrons and photons, which come from the decay or annihilation of the DM particles, in the range of 5 GeV-10 TeV with unprecedented energy resolution (1.5\% at 800 GeV).

DAMPE has also some other scientific objectives which include (1) understanding the mechanisms of particle acceleration operating in celestial sources and the propagation of cosmic rays in the Galaxy, and (2) studying the high-energy behavior of Active Galaxy Nuclei, Galactic pulsars, Gamma-ray Bursts and other kinds of transients and diffuse $\gamma-$ray emissions.

DAMPE consists of four detectors, a Plastic Scintillator Detector (PSD), a Silicon Tungsten Tracker (STK)~\cite{P.Azzarello}, a BGO electromagnetic calorimeter (BGO ECAL)~\cite{BGO_ECAL} and a NeUtron Detector (NUD). It has been successfully launched into a sun-synchronous orbit at the altitude of 500 km on Dec. 17, 2015, and in this paper, we present a comprehensive overview of the PSD, include the design, the manufacture and the performance obtained from the test results on the ground.

%%%%%%%%%%%%%%%% Main Text Body %%%%%%%%%%%%%%%%%%%%%%%
\section{Requirements for PSD}

The PSD is installed at the top of the satellite, and as the first layer of the DAMPE. The purposes of the PSD is to measure the charge information for incident high-energy particles with charge number Z from 1 to 26, which means that the PSD must have high detection efficiency, large dynamic range and good enough energy resolution for charged particles.

The electron/proton separation can be mainly performed by the BGO calorimeter since the horizontal and vertical cascade of electron/proton will appear very different from an electromagnetic shower and a hadron shower, and the separation power will be improved once more through the NUD by rejecting proton. But for $\gamma$ and electron, because they will have the similar behavior in the calorimeter, the identification must be done earlier. To do this, the PSD needs to has less possibility to detect neutralized particles as $\gamma$, which is important for $\gamma$/e separation and also necessary for $\gamma$ astronomy. Therefore, only light materials with smaller thickness can be considered in order to reduce the possibility of electromagnetic shower created by electron and  $\gamma$ within.

Considering the different differential spectra of cosmic ray electrons and photons above 5 GeV in the space, the requirement of misjudgment is that no more than 1\% for electrons and photons separation. As a result, the PSD system is required to provide at least 0.995 efficiency for singly charged Minimum Ionizing Particles(MIPs) with an energy resolution better than 25\%.

According to the design of DAMPE, there will be some tungsten plates inside the STK, and electromagnetic shower created by the incident high-energy electron or photon will has a large possibility occurring close to the PSD. This is so-called back-splash effect, which means that isotropically distributed secondary particles from the shower can hit the PSD and will cause many fake events during the subsequent data analysis. To reduce this effect, an array structure is necessary for the PSD.

As part of a space-borne apparatus, there are also many other constraints on the design and construction of the PSD. Considering the field view requirement of DAMPE, the active area of PSD need to be large than 820 mm $\times$ 820 mm, and this also need to be satisfied with a total weight less than 110 kg and an average power consumption no more than 10 W. Because the PSD system must maintain a high level of performance and stability throughout the whole mission duration, which is at least 3 years in the harsh environment of space, all the chosen components and materials should be satisfied with the radiation hardness requirements, and both of the mechanical and thermal design need special consideration to make sure the PSD can survival from the vibrations, shocks and accelerations during the launch and work well in a broad temperature range of -20 $^{\circ}$C to +45 $^{\circ}$C (storage) or -10 $^{\circ}$C to +30 $^{\circ}$C (operation).

\section{Design of the PSD}

\subsection{Detector description}

Since the photon/electron identification requires that only light materials with smaller thickness can be used in the active area of PSD. The organic plastic scintillator is finally chosen as the detection material due to its small density, high efficiency for charged particles, good radiation hardness, large volume size available and easier to be processed. Because the plastic scintillator is not a rigid material, the honeycomb plates with the carbon fiber reinforced plastics (CFRP) as the skin will be used to make the main support structure for PSD, in order to resist the harsh environments during the launching process.

\begin{figure}
 \centering
 \includegraphics[width=80mm]{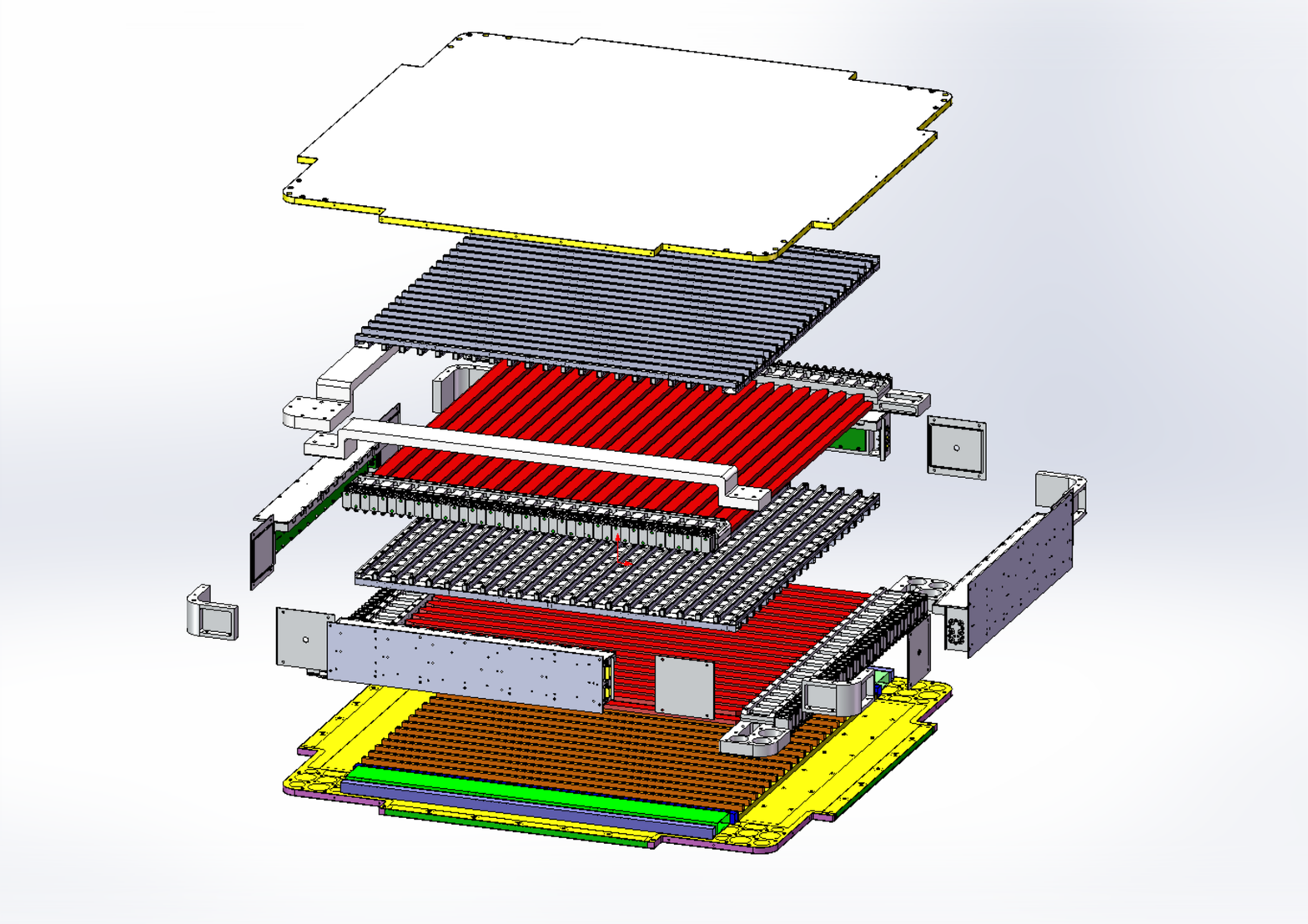}
\caption{The exploded view of the PSD.}
\label{fig:fig1}
\end{figure}

To reduce the influence of the back-splash effect, the PSD is designed to a double layer configuration with 82 detector modules totally, and an overall design of the PSD is sketched showing in Figure \ref{fig:fig1}.

\begin{figure}
 \centering
 \includegraphics[width=80mm]{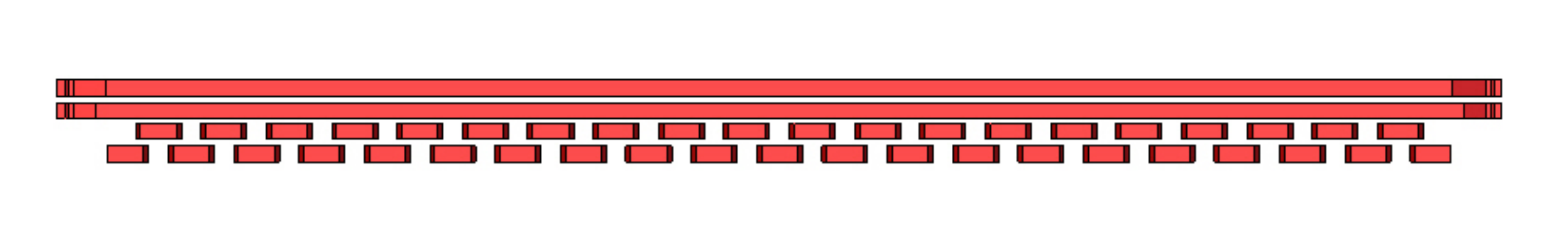}
\caption{The arrangement of the scintillator bars inside the PSD (a side view).}
\label{fig:fig2}
\end{figure}

Each detector module has a long plastic scintillator bar with a dimension of 884 mm $\times$ 28 mm $\times$ 10 mm, and the signals are readout by two Photo Multiplier-Tubes (PMTs) coupled to the ends of the bar. The modules are parallel to each other in one layer, and the directions of the scintillator bars in the two layers are perpendicular. To avoid the presence of any ineffective detection area, as shown in Fig.\ref{fig:fig2}, the neighbored modules in one layer are staggered by 8 mm.

With this crisscross structure, an active area of 825 mm $\times$ 825 mm is fully covered and every incident charged particle will penetrate at least two modules. By this way, a 0.95 detector efficiency for a single module will lead to a higher efficiency of $\geq$ 0.9975 for the whole PSD, and the incident position of the charged particle can also be obtained, which can be coincident with the trajectories derived from the STK and the BGO calorimeter to suppress the noise.

\subsection{Detector Module}
\label{sec:modules}

\begin{figure}
 \centering
 \includegraphics[width=80mm]{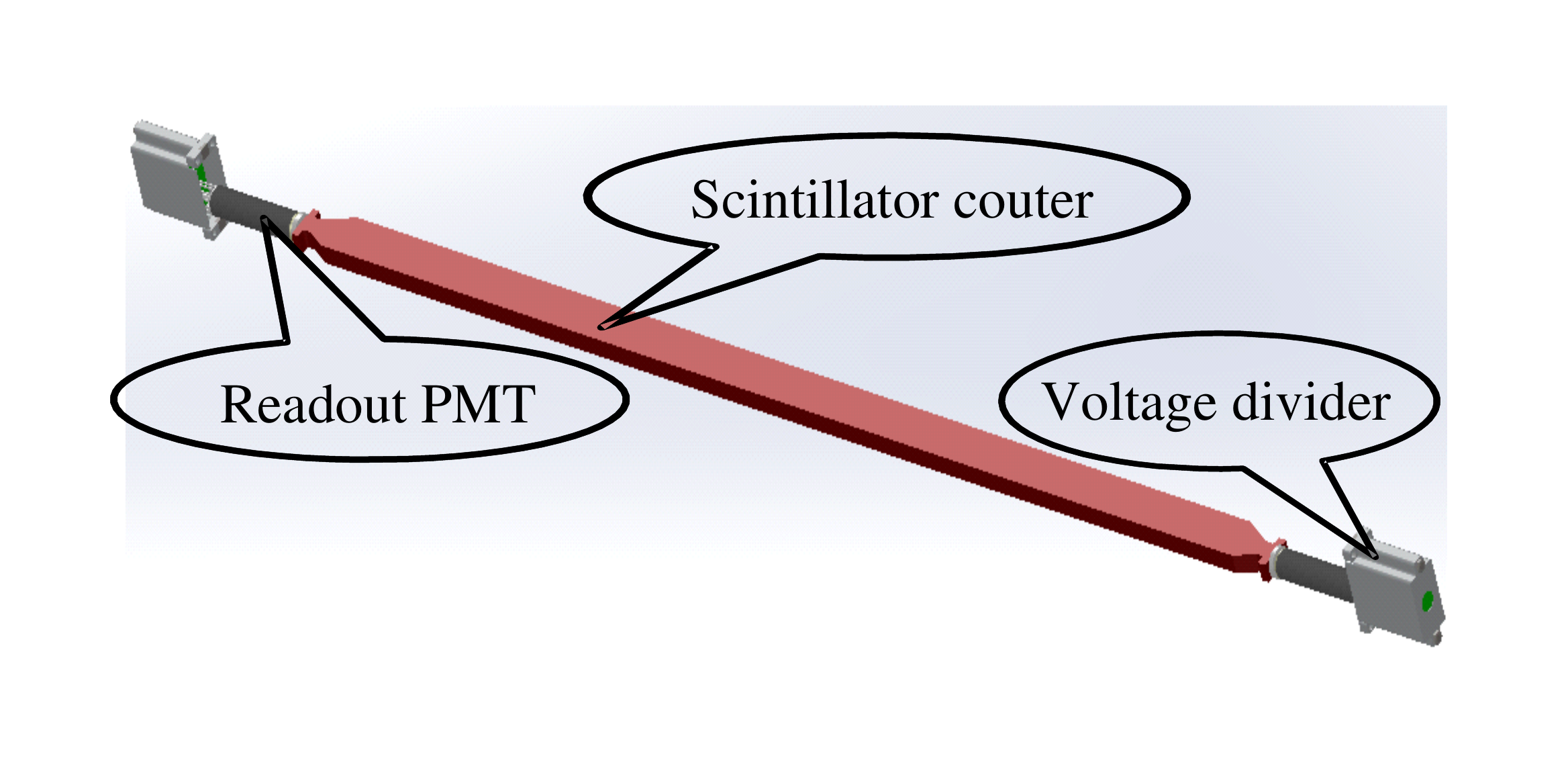}
\caption{A schematic diagram of the PSD detector module.}
\label{fig:fig3}
\end{figure}

Fig.\ref{fig:fig3} shows a schematic view of one detector module, which is mainly composed of two parts: the scintillator bar and the readout devices with the associated circuits.

For the material of the bar, we choose EJ-200 ~\cite{scintillator} produced by Eljen Technology Corporation, which has been used by many space experiences~\cite{ACD_GLAST,AMS_TOF} already. EJ-200 has a good time performance (0.9 ns rise time and 2.1 ns decay time), a relatively higher light output (typically 64\% of Anthracene), a large bulk light attenuation length (380 cm) and the scintillation emission is centered on a wavelength of 425 nm.

The PMTs will be used as the signal readout devices for the modules, and we choose R4443 ~\cite{r4443}, manufactured by Hamamatsu Photonics. R4443 is a 10 stage head-on type PMT with a minimum effective area of $\phi$ 10 mm. It has a low noise Bialkali Photocathode and the quantum efficiency is about 18 \% at 420 nm. This PMT is a ruggedized type and has been already successfully employed in a space-borne experiment ~\cite{ACD_GLAST}, so we think it can meet the environmental related requirements of DAMPE.

Because the cross-section of the bar is smaller, to ensure the mechanical reliability and safety, no separated light guides used to couple the bar to the PMTs. Instead, the end sockets of the bar are special manufactured to act as the light guide, as shown in Fig.\ref{fig:fig3}. And a 3 mm thick optical silicon rubber sheet is used to couple the scintillator bar and the PM photocathode window for reducing the possible damage from vibration and others.

The purpose of each module is to measure and identify the incident particles with charge number from 1 to 26, through their different energy deposition. When charged particles passing through a thin material, they will deposit a certain amount of energy, which is dominated by the ionization process in our case. DAMPE will measure relativistic high-energy particles, and their energy deposition reaches minimum and is nearly independent of the energy if only ionization process is considered. It's normally called  minimum ionizing particles (MIPs) for Z=1 particles like electron. And we mark the energy deposition of these MIPs with a perpendicular incidence as 1 MIP, which is about 2 MeV for the PSD bars. For ions with larger charge number Z, their energy deposition can be described well with Bethe-Bloch formula ~\cite{bethe}, which means that the energy deposition is only proportional to Z$^2$ in the relativistic energy regime.

Consider the incident angle vary and other effects, detailed analysis and simulation shows that even take into account the quenching effect of plastic scintillator, which means the light output for unit energy deposition will be reduced for particles with large charge number and can be well described by Birks-Chou law ~\cite{{dwyer85}}, a dynamic range from 0.1 MIPs to 1400 MIPs is necessary for each PMT, 4 orders in magnitude.

\begin{figure}
 \centering
 \includegraphics[width=80mm]{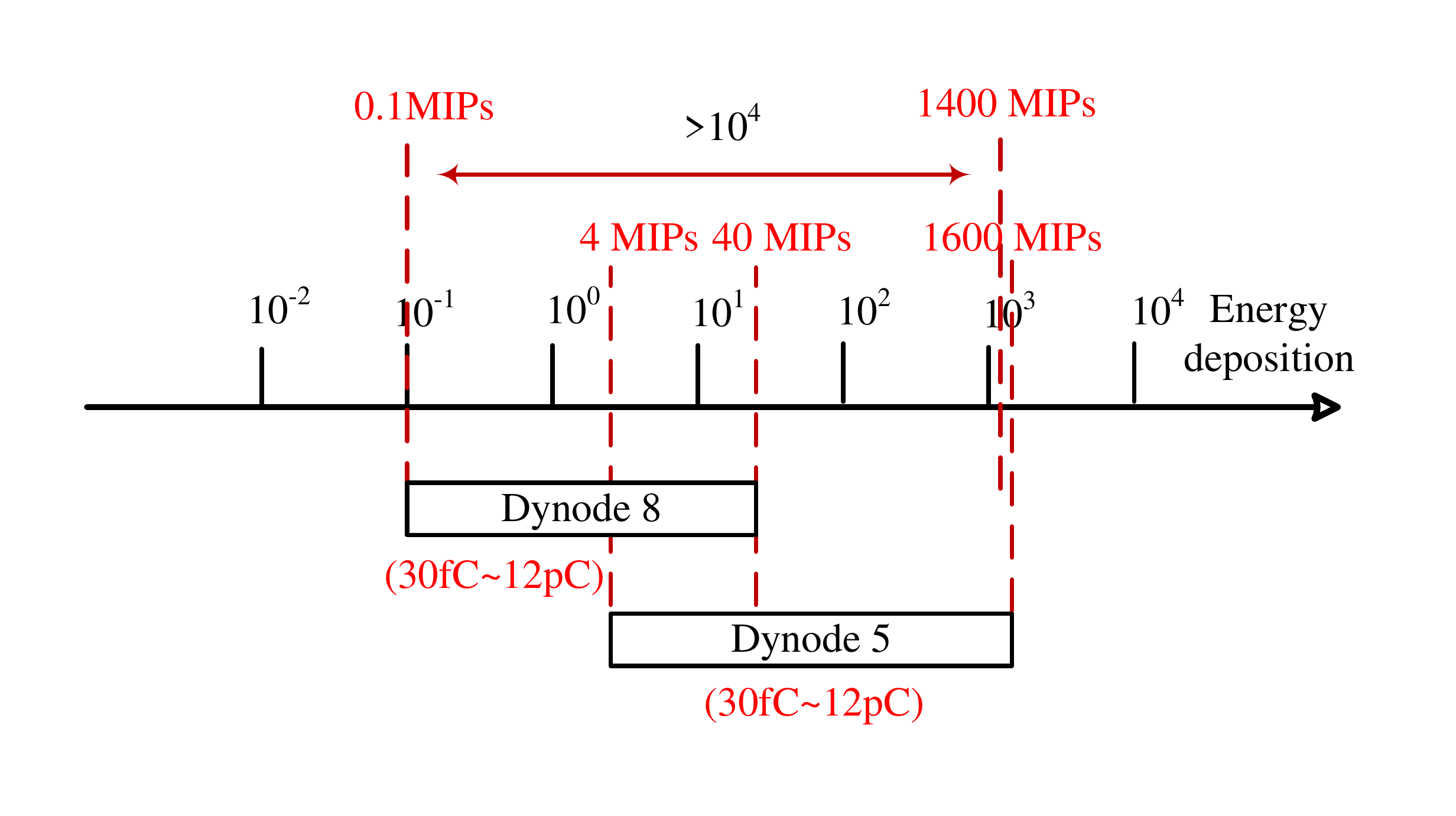}
\caption{The requirement and the design concept of the large dynamic range for the PSD.}
\label{fig:fig4}
\end{figure}

It's very difficult to cover this broad range with good resolution by using only one circuit, therefore we designed a double dynode readout scheme for each PMT, as shown in Fig.\ref{fig:fig4}. Signals from the 5th dynode with smaller gain can cover from 4 MIPs to 1600 MIPs, and the ones from the 8th dynode with bigger gain will cover from 0.1 MIPs to 40 MIPs. And there has some overlap between the signals of two dynodes that can be used for calibration ~\cite{yong}.

\subsection{Electronics}

An array structure is necessary for the PSD, which makes the signal process a little more complicated.
Totally, there are 164 PMTs and 328 signal channels used in the PSD, and they are separated into four groups according to the orientation.

\begin{figure}
 \centering
 \includegraphics[width=80mm]{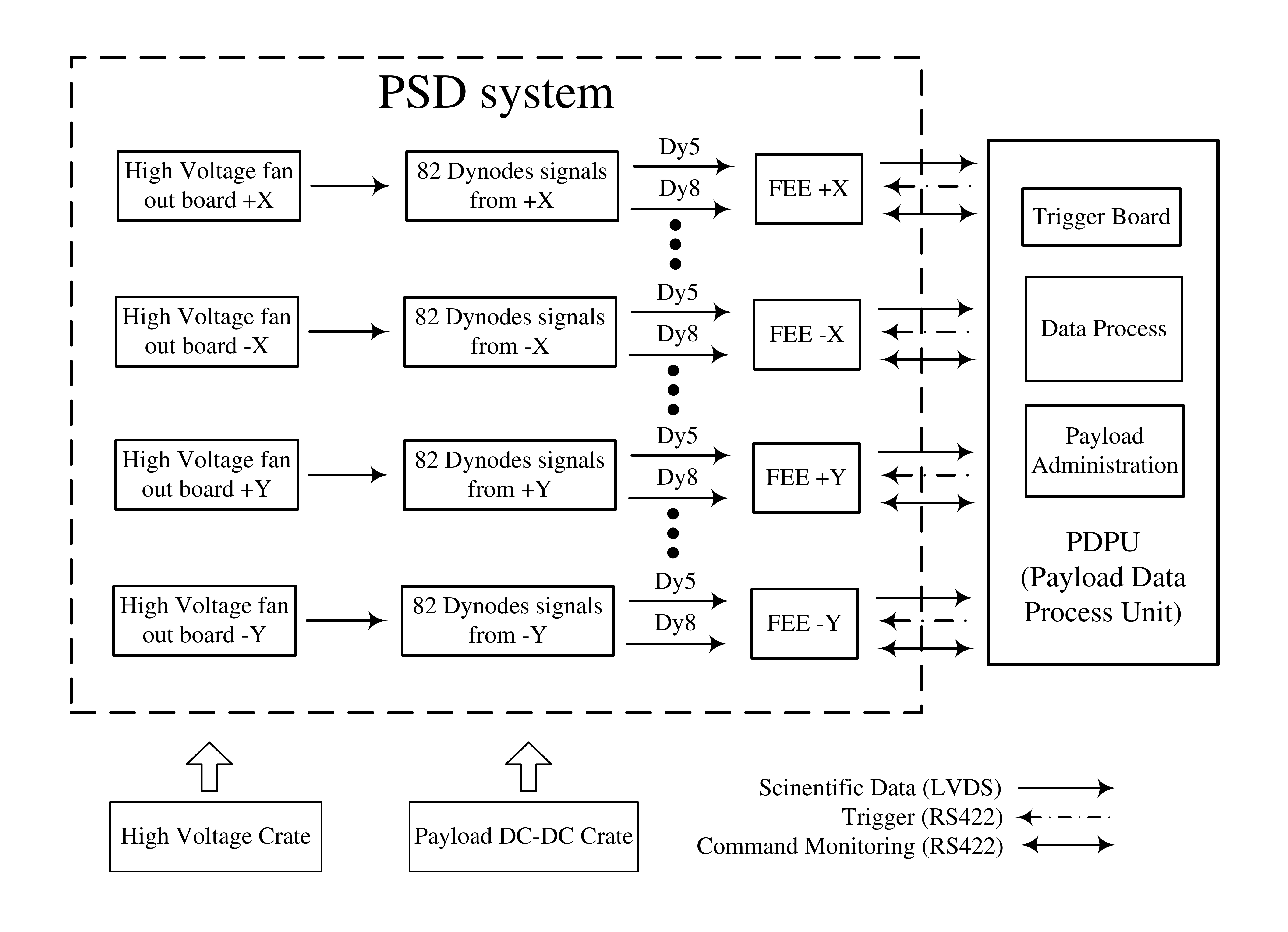}
\caption{Scheme view of the readout electronics for the PSD.}
\label{fig:fig5}
\end{figure}

Fig.\ref{fig:fig5} shows the scheme of the readout electronics for the PSD. In each direction, we have a front-end electronics (FEE) board to handle the 82 signals from 41 PMTs in this side. The coaxial cables from PMTs are soldering on an interface board, which connect with the FEE board with multi-pin connectors from Airborn~\cite{airborn} to get more convenience to the assembly and test works in the future.

Because only charge information is interested for each PSD signal, we design the FEE circuit of PSD based on an ASIC chip VA160, which is a specific modified version of the VA32-HDR14.2, a popular used ASIC chip developed by IDEAS (Norway)~\cite{VA_chip}, to simplify the work. It's a low power consumption ASIC chip and contains 32 channels of charge sensitive preamplifiers, shaping amplifier, and sample hold circuit with a shaping time of 1.8 $\mu$s. For each channel, it has a rms noise level of ~0.8 fC and a 13 pC dynamic range for positive charge inputs.

The PMT dynode signals are coupled to VA160 directly, and each PMT will send its two signals to different VA160 chips to reduce the crosstalk. There are 4 VA160 chips in one FEE board, and the analog to digital conversion is performed by one 14-bit ADC under the control of the onboard FPGA~\cite{fee}.

The data taken of DAMPE works in an event-by-event mode. For PSD, the onboard FPGA of each FEE board always listen for the trigger signal from the DAMPE data process unit, and start the data taken process after checking the validity of this signal. When it's finished, the data from the 41 PMTs in this side will be packed and send to the DAMPE data process unit immediately with a user-defined serial protocol based on LVDS standard.

From a laboratory test, the equivalent noise level for one channel of the PSD is less than 6 fC and the saturation level is about 12 pC. This implies that the response of 0.1 MIP shall be larger than 30 fC while a 5 $\sigma$ separation is required, i.e. 400 ADC channels as shown in Fig.\ref{fig:fig4}.

The FEE board need DC powers like +5.7V, +3.3V to work, and they are provided directly by the DAMPE power supply system. Except this, this system also provide the high-voltages up to 1250V needed by the PSD PMTs to work properly, and in order to reduce the HV cables and modules of the whole DAMPE system, only 6 HV inputs are provided for each side of the PSD. So accompany with each FEE board, there also have a HV fan-out board. On this board, one HV input will be split into 7 (one is 6 because we have 41 PMTs), and for each output, an additional resistor of 3.9 M$\Omega$ is used to limit the current and protect the HV input in case the failure of the PMT or its voltage-divider.

\subsection{Mechanical and thermal design}

The PSD has an active volume of more than 825 mm $\times$ 825 mm but only 40 mm in thickness, which means the whole structure is relative fragile for a space-borne equipment. And this becomes even worse while no mechanical anchor point allowed within the active area and we have a weight limitation.

To solve this problem, the CFRP are used as the main structure material. The base board, which is the primary force bearing plate of the PSD, is a large  10 mm thick honeycomb plane with 1 mm CFRP skin in both sides, and some CFRP cubic hollow tubes are buried into the honeycomb to enhance the strength. The 82 scintillator bars are arranged into four layers, every bar is restricted in a groove, which are formed by CFRP made cubic hollow tubes glued in the inner-surface of the base board or the CFRP made dummy plates between the layers.

Surround this active volume, there are four crossbeams to house and protect the PMTs for the detector modules, and these crossbeams are made with aluminum due to its much better machinability than the CFRP.

\begin{figure}
 \centering
 \includegraphics[width=70mm]{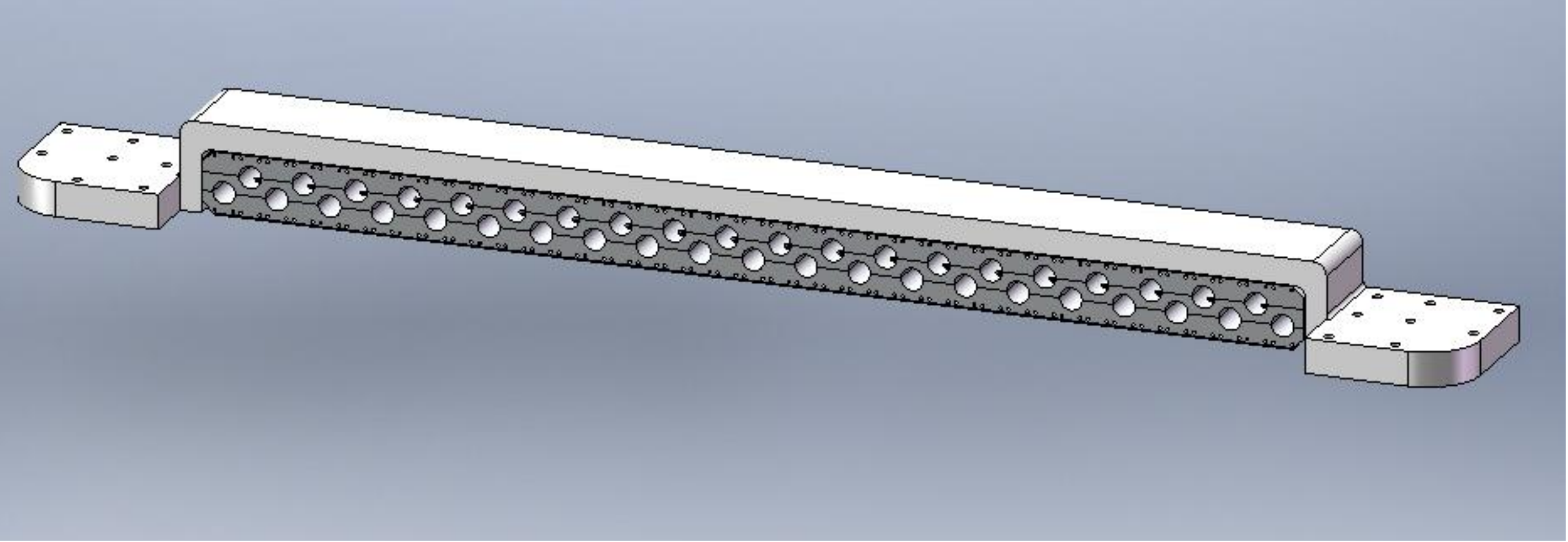}
\caption{The CAD drawing of the aluminum crossbeam and the CFRP U-shape clamp.}
\label{fig:fig6}
\end{figure}

Because there is large difference in the temperature coefficient between the plastic scintillator (7.8 $\times$ $10^{-5}$ / $^{\circ}$C) and the CFRP (zero or even less), consider that we have a 884 mm long bar and a large temperature variation range of more than 50 $^{\circ}$C, the length of the detector modules may change up to 4 mm in different temperature while the size of the base board is not changed, and this difference may cause damage to the modules. In order to avoid it, for each layer, all the modules are only fixed with the PSD base board through the crossbeam in one side, and the crossbeam in another side is only constraint by a U-shape clamp and keep the freedom along the modules direction, as shown in Fig.\ref{fig:fig6}. This U-shape clamp is made with CFRP also and special slices with smaller friction coefficient are glued between the clamp and the crossbeam to reduce the resistance.

\begin{figure}
 \centering
 \includegraphics[width=70mm]{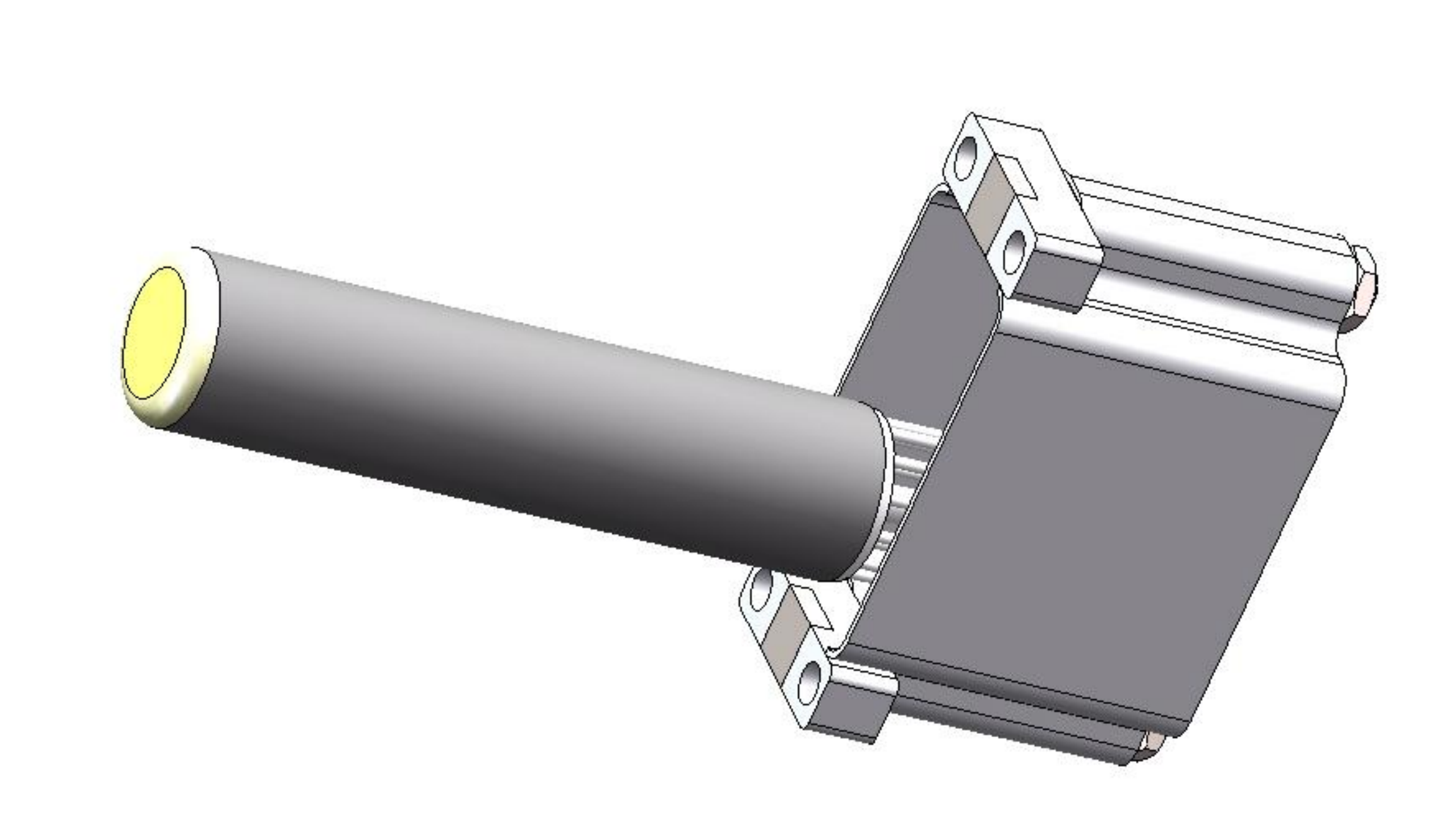}
\caption{The layout of a PMT module.}
\label{fig:fig7}
\end{figure}

The crossbeams (clamps) are stacked at the four corners of PSD, which are also the anchor points for connecting with the STK, and tighten with the base board as a whole. There are many holes in the crossbeam as shown in Fig.\ref{fig:fig6}, and the PMTs of the detector modules are placed in these holes. Fig.\ref{fig:fig7} shows the layout of a PMT module, and the PMT pins and the associated PCBs are housing in an aluminum made box for electromagnetic shielding. Through this box, the PMT is fixed with the crossbeam both mechanically and thermally.

In each side of the PSD, as part of the sidewall, we have an aluminum made box for the FEE board, which can give good electromagnetic shielding and thermal conducting for the FEE board.

The top cap is also a 10 mm thick honeycomb plane with 1 mm CFRP skin in both sides. It connected to the sidewalls and floating above the active part of the PSD.

To assure the reliability of the design, a detailed finite element analyzing has been performed for the system. The results shows that our design has enough safety factor and either the stress or the deformation anywhere are much smaller than the limitation of the materials used. For example, the modal analysis shows that the first order modal frequency is around 128.4 Hz, much higher than the 70 Hz required by the satellite.

Because the output of PMT is dependent on the temperature, a stable thermal environment avails the better performance of the PSD. There are four FEE boards and 164 PMTs, which are the only thermal source in the PSD. The power consumption is about 1.5 W for each FEE board, and less than 15 mW for one PMT. So the total power dissipation is no more than 8.5 W and distributes in the fringe area only. These heat dissipations are transferred to the sidewalls through the aluminum made components, and thermotube are used to connect the four sidewalls together to get a more homogeneous temperature distribution. With this design, the finite element analyzing shows that the maximum temperature difference to the radiating surface is less than 3 $^{\circ}$C for the PMTs and 6 $^{\circ}$C for the FEE boards.

The DAMPE plans to work in a sun-synchronous orbit and one of the sidewall can never have the sun shining. We use this side as the heat radiating surface and cover all the other surfaces of the PSD with thermal insulation foils. In order to keep the temperature of this radiating surface stable, an active temperature control strategy is adopted by the satellite with the help of the thermal sensors and heater bands in the surface.

\section{Manufacturing and Assembly}

\subsection{Plastic Scintillator Bar}

After delivered from the company, the plastic scintillator bars were firstly inspected with magnifier to insure that no visual damage had taken place during the shipment. And their dimensions and the temperature coefficients were also checked.

To improve the light collection efficiency and uniformity, the body part of every scintillator bar is wrapped with Tyvek paper produced by Dupont ~\cite{tyvek} for light reflection, and then covered with a black Polyvinyl chloride tube for tightness and light isolation. And due to their irregular shapes, the two ends are only wrapped with Teflon tapes.

\begin{figure}
 \centering
 \includegraphics[width=80mm]{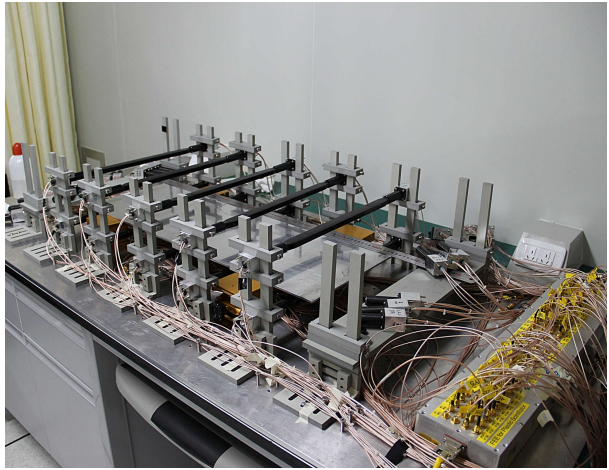}
\caption{The test bench for performance check of the scintillator bars. }
\label{fig:fig8}
\end{figure}

Some performance, such as the light output, the light attenuation length, need to be measured for every bar before it can be used. And due to the narrow shape of the bar, the attenuation length of the light response as quoted by manufacture is not appropriate because it’s only valid for a bulk material. So we use the technical attenuation length to describe the light output uniformity instead of the bulk attenuation length.

For this purpose, a customized test platform has been adopted to check the performance of these scintillator bars by using the cosmic rays ~\cite{zhang2015}, as shown in Fig.~\ref{fig:fig8}. For each bar, information like the light output at the center of the bar, the technical light attenuation length, the detection efficiency and the energy resolution as functions of the position along the bar were obtained and used for selecting the appropriate ones for the PSD.

\begin{figure}
 \centering
 \includegraphics[width=80mm]{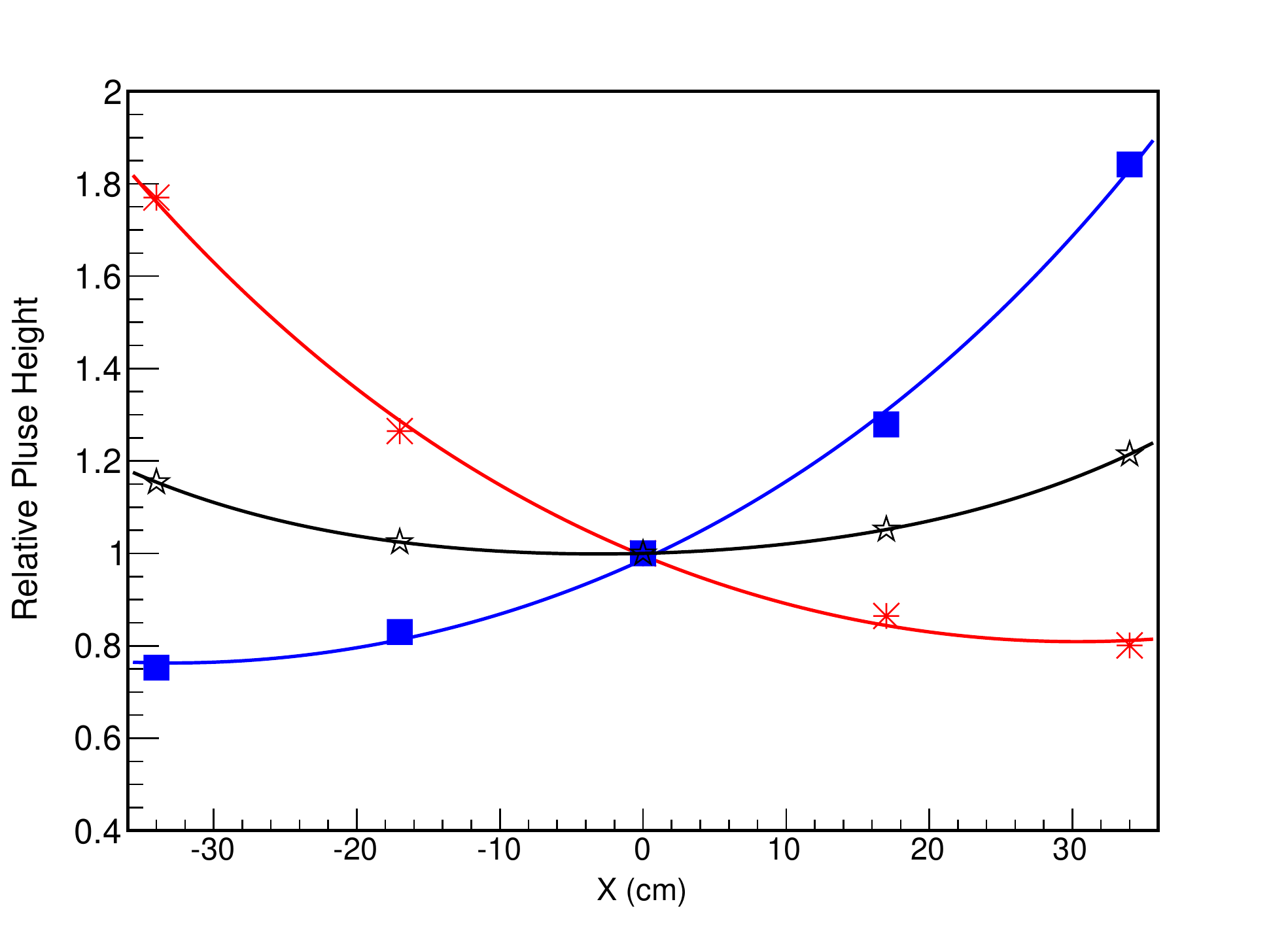}
\caption{The relative light output along the scintillator bar.}
\label{fig:fig9}
\end{figure}

Fig.~\ref{fig:fig9} shows a typical result for the light output, which is normalized at the center of the bar, changed while the impact position of the incident particle varied along the bar. The red and blue points are measured at different ends of the bar, and the black one is the average of these two. The result can be described precisely by the model~\cite{taiuti}:
\begin{equation}
A(x)=C_0(e^{-x/\lambda} + \alpha e^{(2L-x)/\lambda})
\end{equation}
here L is the length of the bar and $\lambda$ is the technical attenuation length. This model considered also the photons reflected back from the end far from the measurement, and $\alpha$ is an empirical factor stand for this contribution.

With this model, the technical light attenuation length can be achieved by fitting the data obtained at various positions. This work has been done for all the bar, and only the ones with the technical light attenuation length larger than 70 cm for both ends can be used in the PSD.

Finally, 82 bars are selected from 150, and except the attenuation length, good detector efficiency, which means better than 95\% at all measurement positions, and large enough light output at the center of the bar are also considered.

\subsection{PMT Module}

Because the harsh environment the PSD may face during launch, the PMTs used would have to be vibration tested to at least the required qualification test levels and critically inspected/tested afterwards. Because it may cause damage to the PMTs, this additional test can only be offered on a best effort basis. So we required the Hamamatsu company to randomly select 5\% of the PMTs from a batch of production and do the vibration test.

A long lifetime is also required for the PMTs because DAMPE will work in space for at least 3 years. So we also asked the company to perform a 12\% hour burn-in test (the so called lifetime test) for every PMT and only the ones with gain difference less than 25\% after the test can be delivered to us.

After the delivery, visual and dimensional inspections were performed to select the intact PMTs. And before soldering with the voltage divider boards, some key performance parameters like the relative gain of dynode 8 and the gain ratio between dynode 8 and dynode 5 at different working voltages were measured by a LED test by using a versatile PMT test bench ~\cite{zhou2016}.

\begin{figure}
 \centering
 \includegraphics[width=80mm]{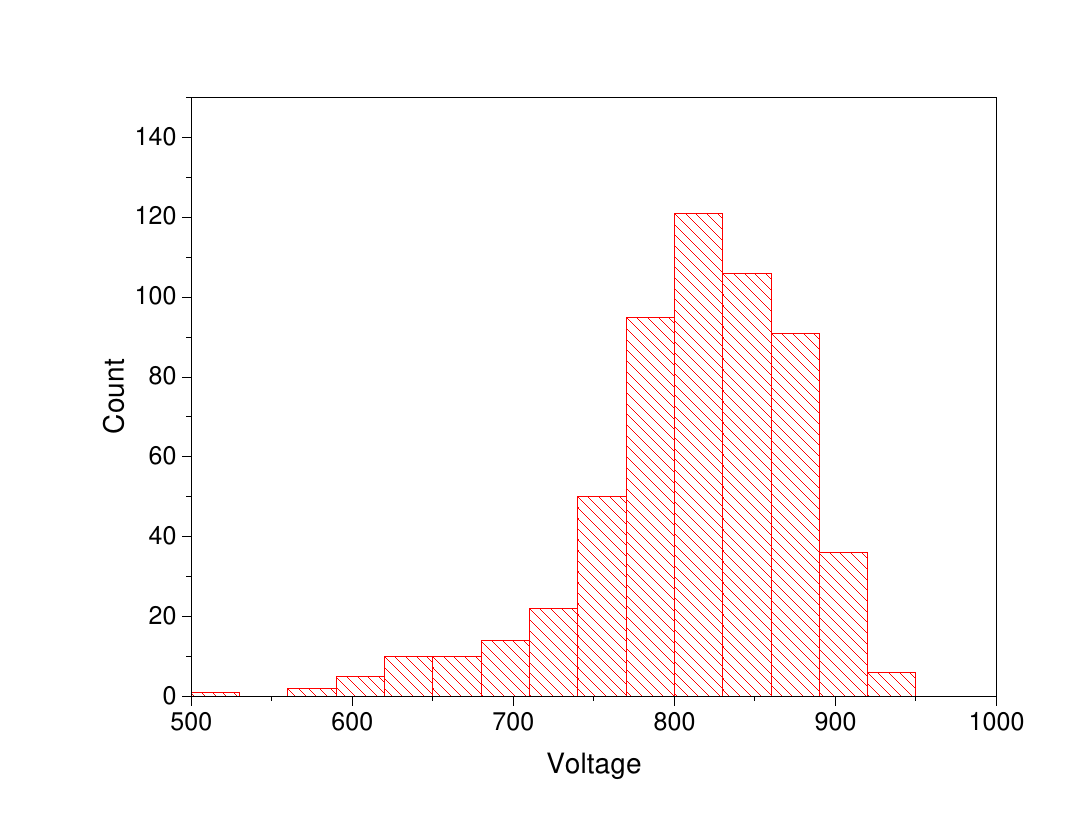}
\caption{The working voltage distribution of the PMTs for a certain gain. }
\label{fig:fig10}
\end{figure}

For a certain signal amplitude, Fig.~\ref{fig:fig10} shows the working voltages needed by the PMTs we tested, and we can see a broad distribution. According to the design, 6 or 7 PMTs will share the same HV input, so we have some uniformity requirement for the gain of the PMTs. From this test, every PMT will has a suggested working voltage, and for the ones with extreme values will be discarded.

\begin{figure}[!ht]
	\centering
	\includegraphics[width=40mm]{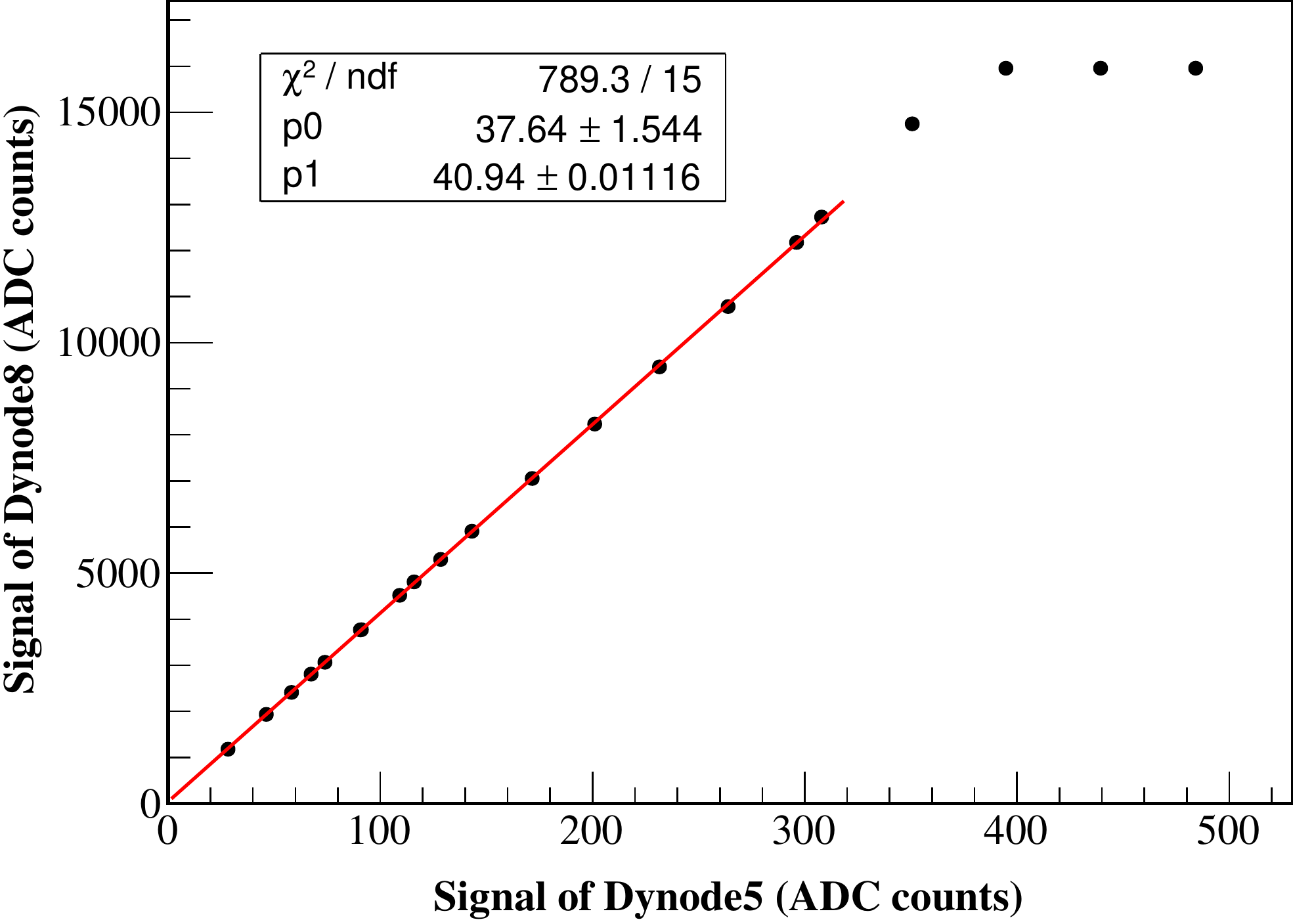}
	\includegraphics[width=40mm]{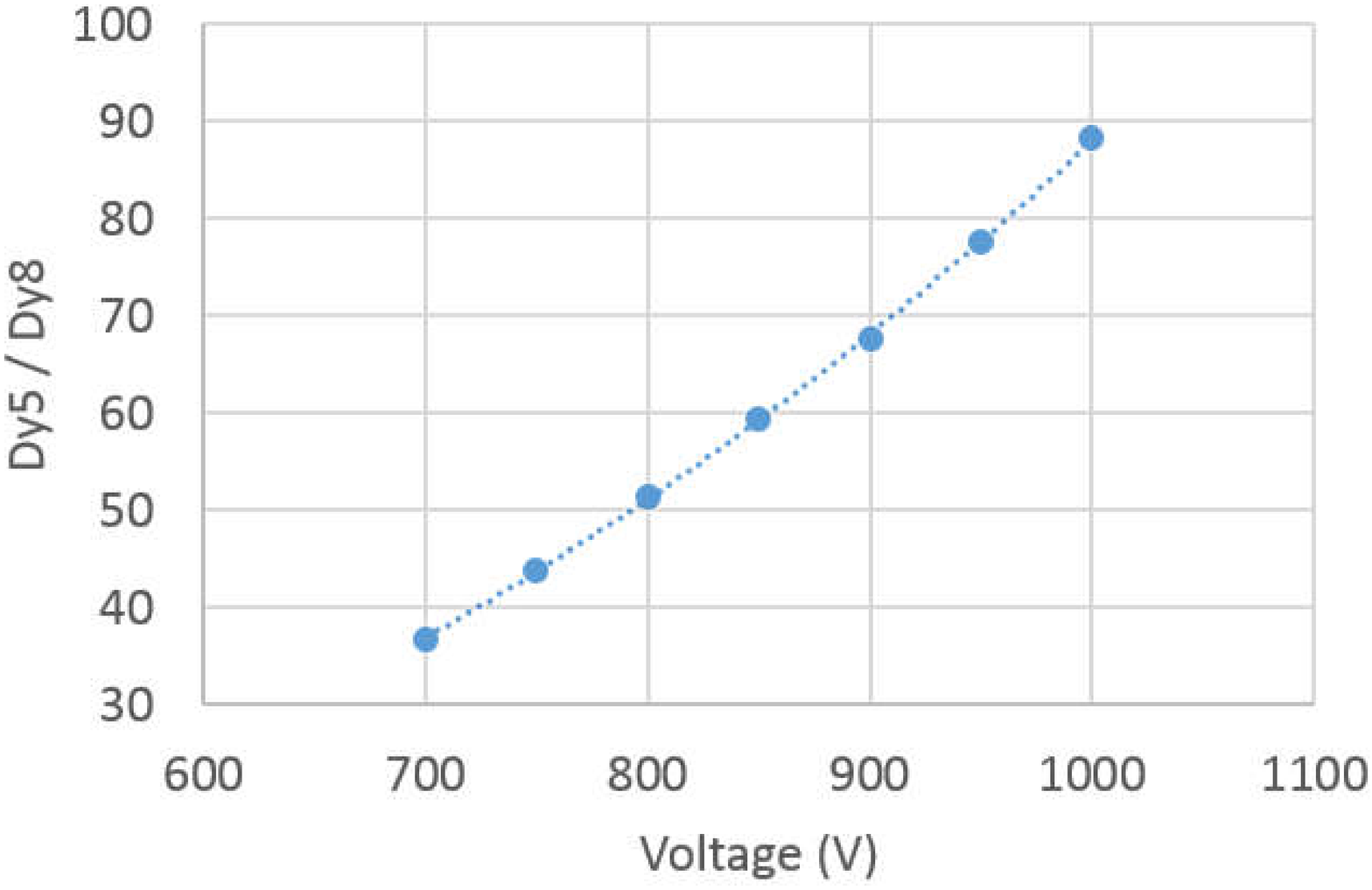}
	\caption{(a)A typical relationship between the double readout dynodes from a LED test.(b)The relation between the ratio and the working voltage.}
    \label{fig:fig11}
\end{figure}

Fig.~\ref{fig:fig11}a shows the relevance between the two dynode signals of the same PMT, which we can get the gain ratio between Dynode 5 and 8 by a linear fit. Fig.~\ref{fig:fig11}b shows the dependence of this ratio on the working voltage for one PMT, which can be used to calculate the ratio corresponds to its suggested working voltage. And only the ones with this ratio larger than 40 at the suggested working voltage can be used in the PSD because the dynamic range we need.

With these conditions, 180 PMTs have been selected for further production, and an initial working voltage was determined for each selected candidate PMTs.

%%\begin{figure}[!ht]
%%	\centering
%%	\subfigure[]{\label{subfig:fig11a} \includegraphics[width=40mm]{fig11a}}
%%	\subfigure[]{\label{subfig:fig11b} \includegraphics[width=40mm]{fig11b}}
%%	\caption{(a)A typical relationship between the double readout dynodes from a LED test.(b)The relation between the ratio and the working voltage.} \label{fig:dy58r}
%%\end{figure}

For the PMT voltage divider, a cathode grounding circuit has been developed and a uniform voltage distribution ratio (1:1:1:1:1:1:1:1:1) is because we are more interested in the linearity than the time performance. The working current is designed to be around 15 $\mu$A because the desired count rate for each module is several tens Hz only. A parallel capacitor filtering network is adopted for the last six dynodes to keep the inter-stage voltages stable and assure a good linearity. The coaxial cables are used for both high voltage and the dynode signals, and for robustness reason, two serial-chained capacitors instead one are used for isolating the HV to the ground and the two readout dynodes.

There are 41 PMTs in one side of the PSD and distance of adjacent ones are only 2 cm. Due to this dense arrangement, the area of the PCB boards for the voltage divider must be less than 2 cm $\times$ 3.5 cm, and only one board is not enough to contain so much components.

\begin{figure}
 \centering
 \includegraphics[width=70mm]{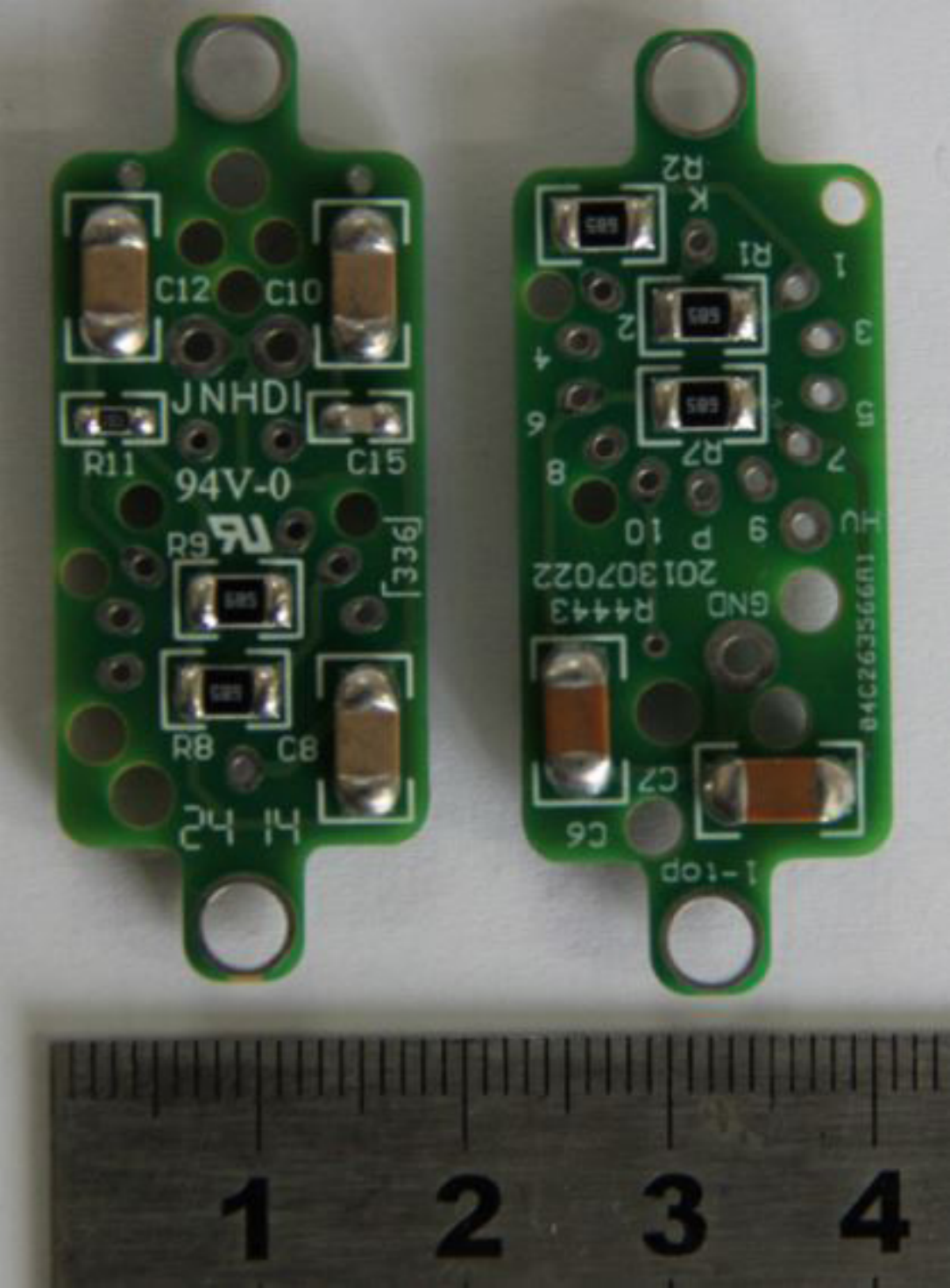}
\caption{The PCBs compose a PMT voltage divider.}
\label{fig:fig12}
\end{figure}

Fig.\ref{fig:fig12} shows the PCBs for one voltage divider, which consist of two 2 mm thick PCBs and more than 20 components. Each board has four copper layers, the top and the bottom layers are used for routing the components, while the two inner layers are used for conducting the heat to the crossbeams of the PSD.

After soldering with the voltage divider boards, the PMT body is potted with a black RTV silicone (RTV627) produced by Momentive Corporation~\cite{rtv} for light screening, and it will also act as an elastic cushion between the tube and the wall of the crossbeam to reduce the risk of damage due to the shocks and vibrations in the launch phase. The PMT pins and the associated PCBs are also potted with a transparent RTV silicone (RTV615), which is produced by the same Corporation, to protect them against the shocks and vibrations as well as the low pressure discharges.

\begin{figure}
 \centering
 \includegraphics[width=80mm]{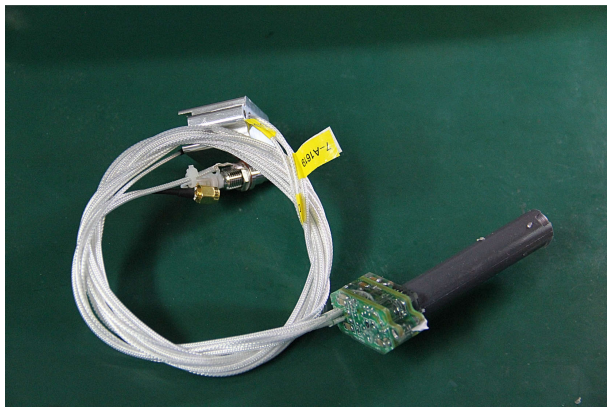}
\caption{A PMT module after potting.}
\label{fig:fig13}
\end{figure}

Fig.\ref{fig:fig13} shows a potted PMT module. After a visual inspection for assuring that no gas bubbles existing in the potted part, latter half of the PMT is covered with the aluminum shielding box, and followed with some tests include the sparking test at a pressure around 1 Pa and a HV value of 1000 V, the temperature test cycling from -15 $^{\circ}$C to +35$^{\circ}$C and the performance verification with LED.

\subsection{Assembly}

For the assembly of the PSD, a custom-made platform was made to simulate the fitting surface in the STK, and
the bottom plate of PSD was fixed on this platform through 16 screws distributed evenly at 4 corners as the real case in DAMPE.

The PSD has 82 individual detector modules. Each module consists of a plastic scintillator bar and 2 PMT modules, and both have contribution to the signal outputs. According to the requirement, the difference of the outputs need to be no more than  25 \%. So the PMTs and the bars need to be matchup carefully. The match principle is to make the product of the PMT relative gain and the bar relative light output as close as possible for all the detector modules. And this was done after the selection of the scintillator bars and the PMT modules.
The position of each detector module in the PSD is also decided by considering that 6 or 7 PMTs will share the same HV power and their performance must be comparable.

According to the design, there are two layers of scintillator bars in one direction, and the ends of the bars are restricted inside the crossbeams, so each crossbeam is divided into three layers for an easy assemble.
The bottom layers of the crossbeams in both sides were placed or fixed in the desired positions on the base board firstly, then based on the matchup results, the bars in the lowest layer were placed into the corresponding grooves on the base board, and restricted by fixing the interlayers of the crossbeams. After this, we put the second layer of bars on the convexes, which formed the grooves, and covered with the dummy plate and the top layers of the crossbeams. The dummy plate was fixed on the base board by many screws evenly distributed along the edge, and the gaps of the grooves are filled with thin poron slices for vibration absorption.

Similar things were done for another direction, and all the detector modules, the crossbeams are tighten together with the base board. The PMT modules were inserted into the holes in the four crossbeams according to the module matchup, and the body part of the PMT was covered with a 60 $\mu$m thick $\mu$-metal foil to prevent the influence of the geomagnetic field in the orbit.

A cosmic ray test was performed to assure the quality of the installations before routing and soldering the cables, and then, all the coaxial cables were routed and soldered on the interface boards, which adopted a pseudo coaxial cable design to reduce the crosstalk, and the HV fan-out board in the certain order. The free part of the cables were banding together or adhering in the crossbeam to reduce the vibrations during launch, and all the welding spots were coated with Momentive RTV118~\cite{rtv} for protection.

\begin{figure}[!ht]
	\centering
	 \includegraphics[width=80mm]{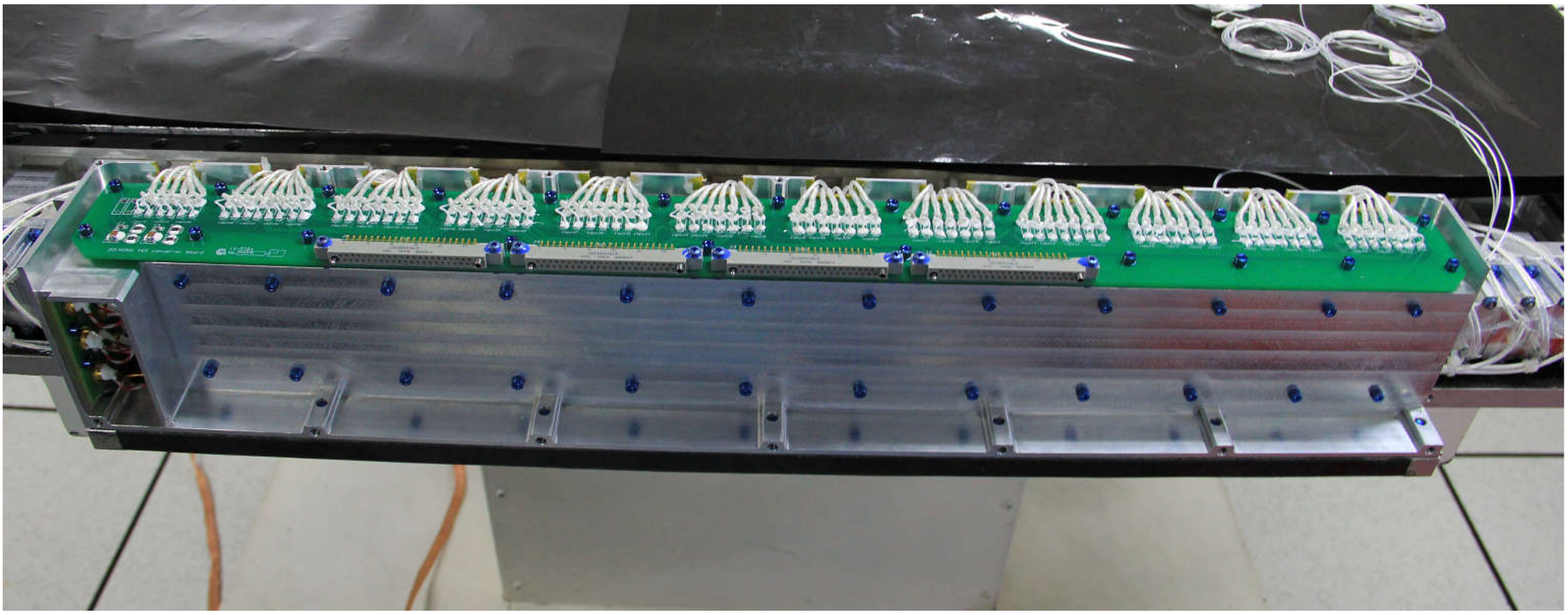}
	 \includegraphics[width=80mm]{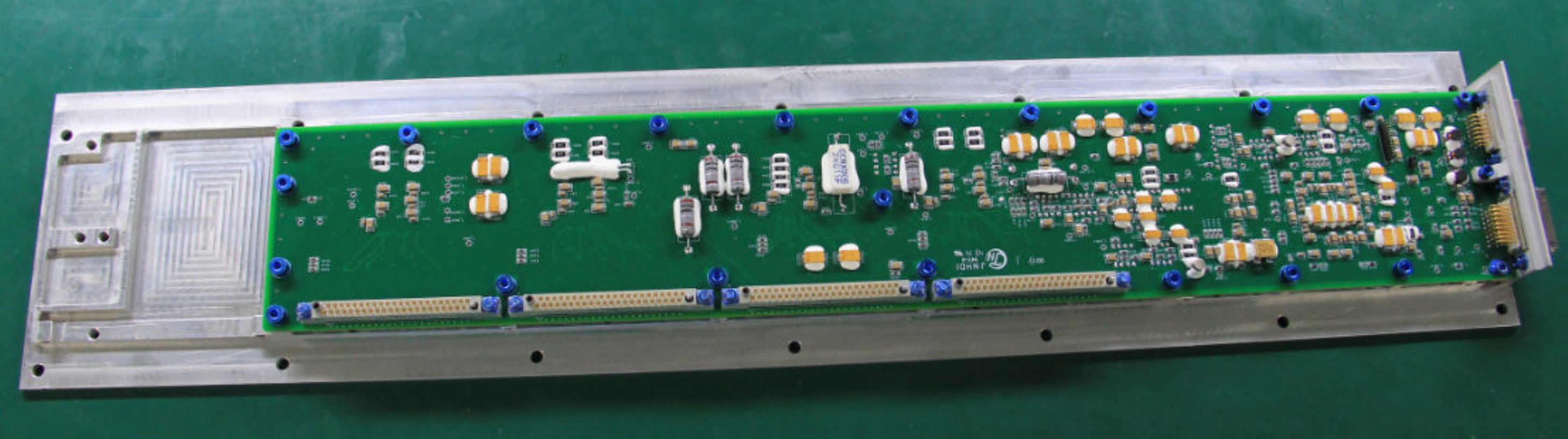}
	\caption{(a)The side view of the PSD just after the cabling and mounting the FEE box. (b)The FEE board mounted on a sidewall board of the PSD.}
      \label{fig:fig14}
\end{figure}

In each side of the PSD, the interface board and the HV fan-out board were mounted on the aluminum made FEE box, and this box was fixed on the bottom plate, as shown in Fig.~\ref{fig:fig14}a. Fig.~\ref{fig:fig14}b shows the FEE board mounted in the sidewall of the PSD, this board can be fit well to the FEE box and fastened with screws.

\begin{figure}
 \centering
 \includegraphics[width=80mm]{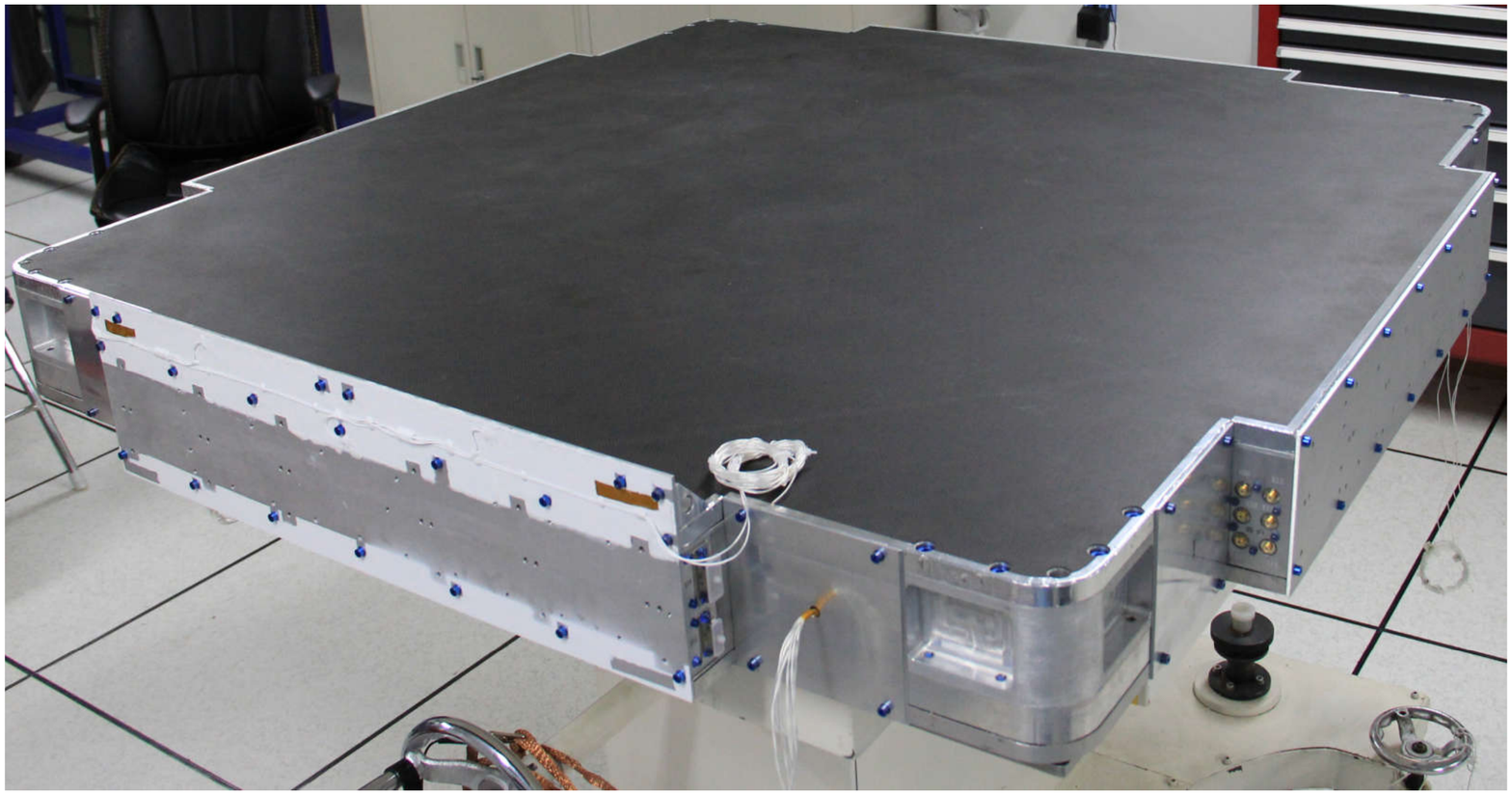}
\caption{The totally assembled PSD flight model.}
\label{fig:fig15}
\end{figure}

The last step of the assembly is to cover the top cap, and Fig.~\ref{fig:fig15} shows the fully assembled flight model of the PSD, which has a final weight of 103 kg and power consumption of 8.5 W.

There have two models been assembled based on the same procedure. One is the flight model (FM), which will be used in the real mission, and another is the engineer qualification model (EQM), which will be used to verify the reliability of the design and assembly works.

\section{Performance and Qualification Test of the PSD}

After the assembly, the performances of the PSD have been checked by a test using the cosmic ray, which mainly consists of high energy muon and the behavior in PSD is similar to the MIPs.

\begin{figure}
 \centering
 \includegraphics[width=80mm]{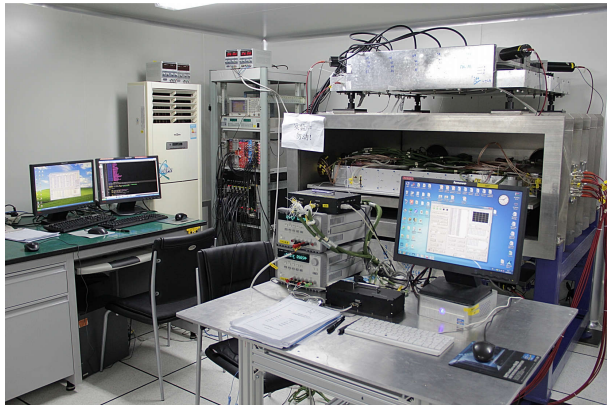}
\caption{The platform for testing PSD with cosmic rays. }
\label{fig:fig16}
\end{figure}

As shown in Fig~\ref{fig:fig16}, the PSD was housed in a custom-made platform in the test, and there have two sets of auxiliary detectors above and below it. Each set of auxiliary detectors consist of a plastic scintillator detector for triggering and a multi-wire drift chamber (MWDC) for position measurement, both have active area a litter larger than the PSD. All these auxiliary detectors were synchronous with the PSD, and the whole apparatus was triggered by the coincidence of the two large trigger detectors for an incident cosmic ray particle. The trajectory of the particle was obtained with the help of the MWDCs, and it would be used in determining the hit position in the PSD and applying the angle correction, which will eliminate the influence of material thickness vary induced by the change of incident angle.

\begin{figure}[!ht]
	\centering
	\includegraphics[width=40mm]{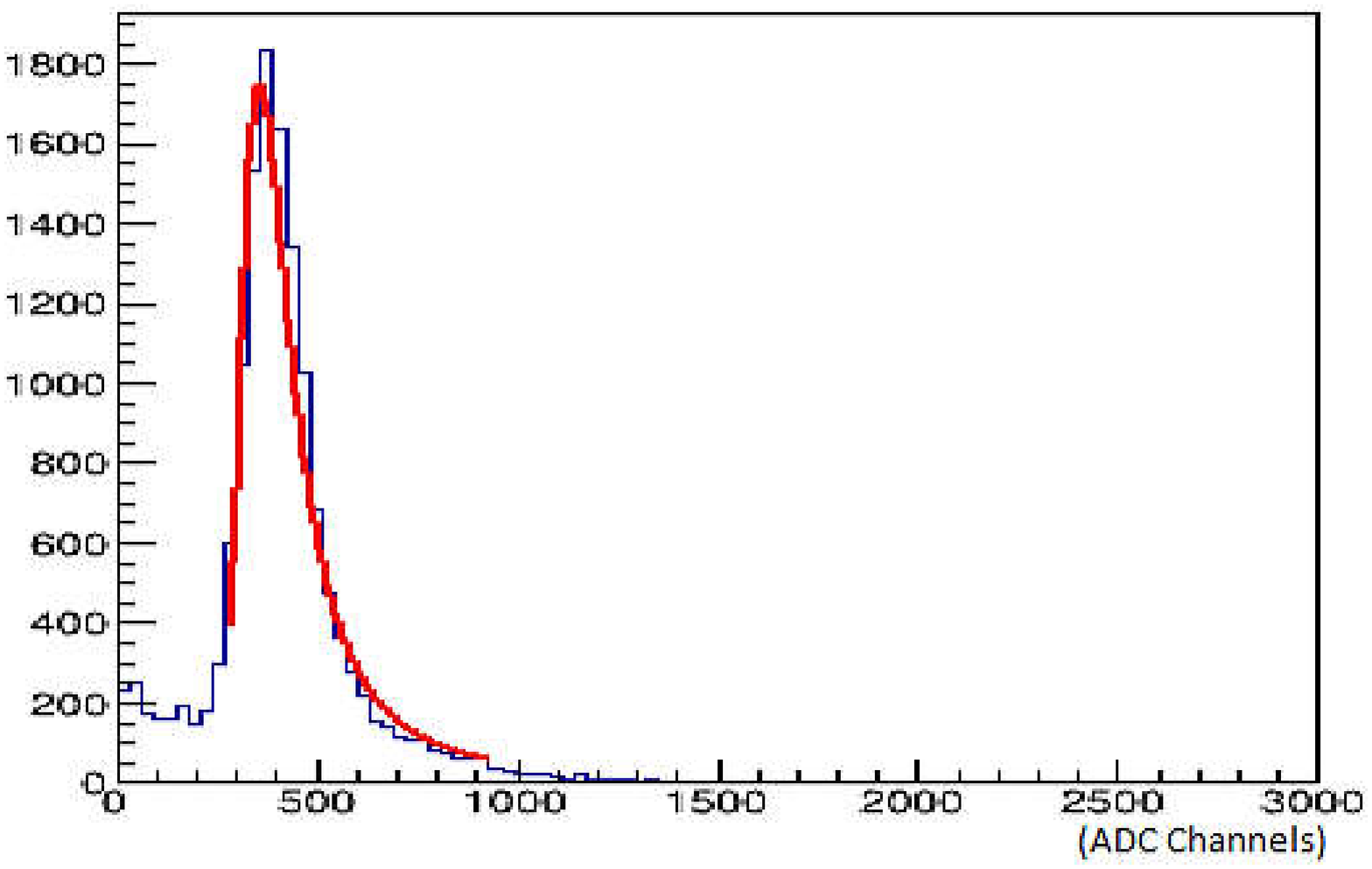}
	\includegraphics[width=40mm]{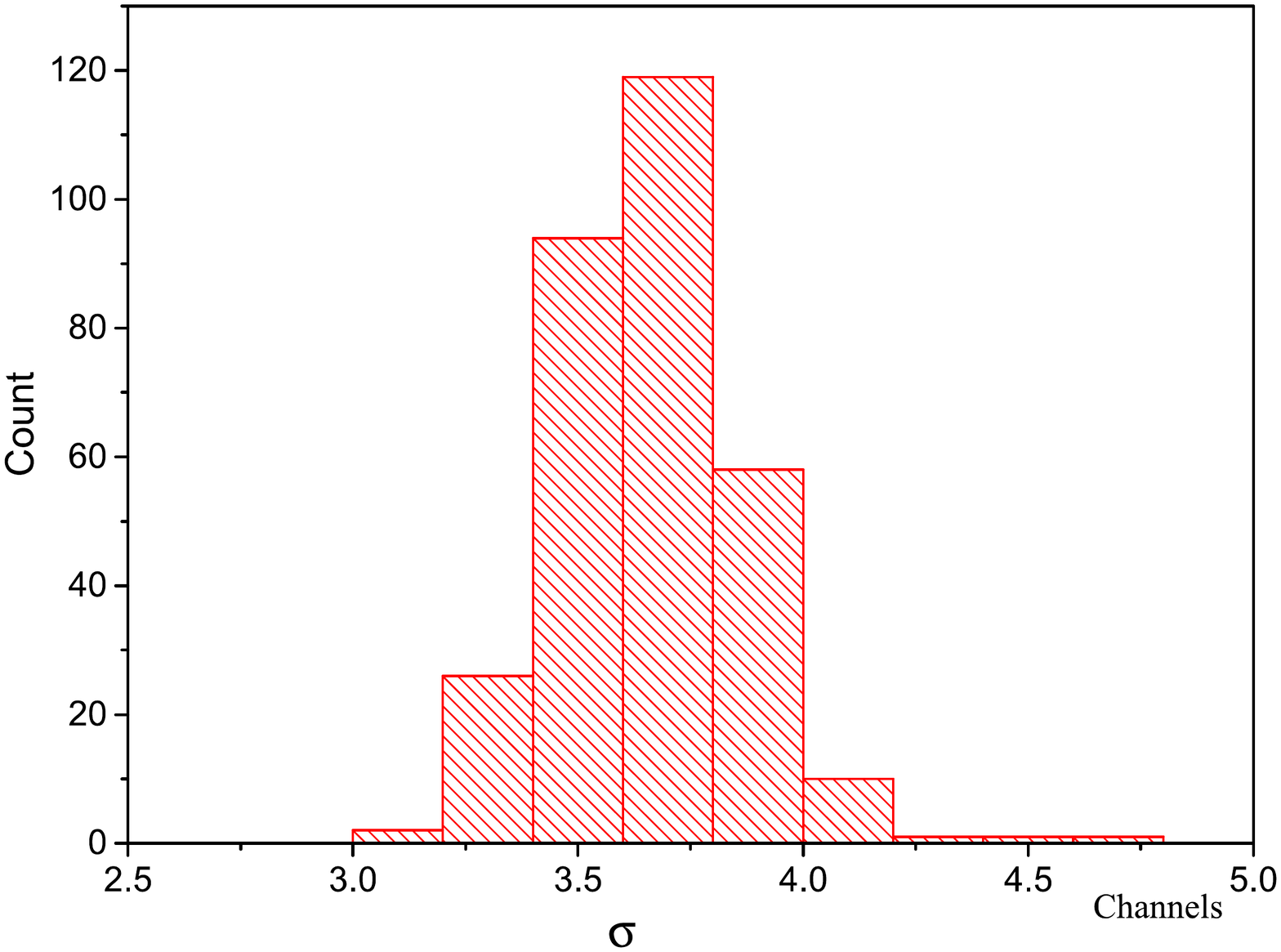}
	\caption{(a)The pulse height distribution from the 8th dynode of a PMT for the incident cosmic rays.(b)The width ($\sigma$) of noise spectrum distribution for the PMTs used by the PSD.} \label{fig:fig17}
\end{figure}

The spectrum obtained from the cosmic ray test is shown in Fig.\ref{fig:fig17}a for one PMT. It can be well described with the landau function, and we can obtain the most probable value (MPV) and others by fitting the spectrum. The noise level for the detector was measured by using a periodic signal from a signal generator as the trigger, and the obtained noise spectrum can be fit well with a Gaussian function. Fig.\ref{fig:fig17}b shows the distribution of the width($\sigma$) of the noise for all the PMTs, and we use 5$\sigma$ as the threshold in the further analysis. From Fig.\ref{fig:fig17}a, we can see a really good signal over noise ratio and the MPV is more than 10 times of our identify threshold.

\begin{figure}
 \centering
 \includegraphics[width=80mm]{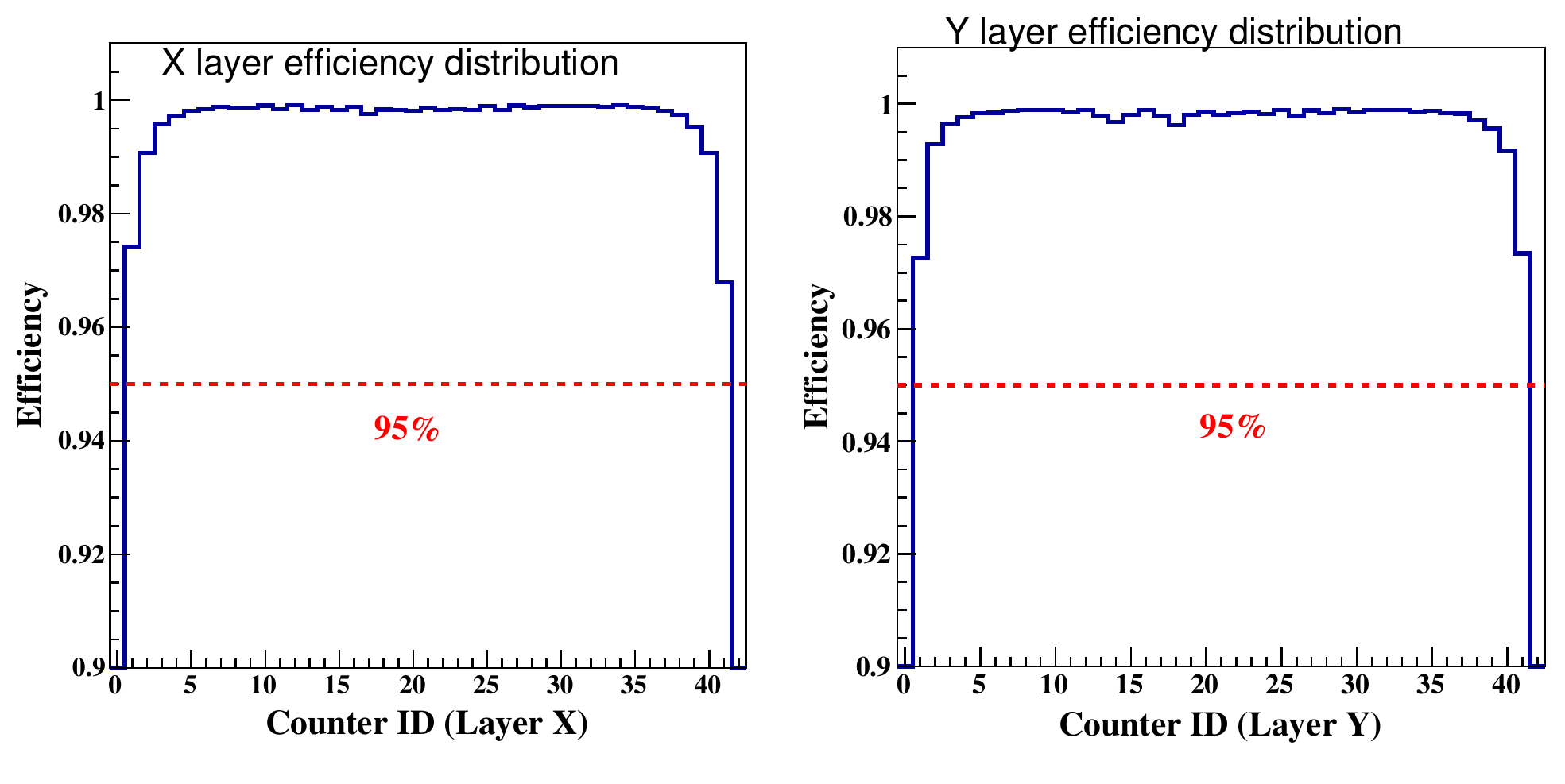}
 \caption{The detection efficiency for all the detector modules.}
 \label{fig:fig18}
\end{figure}

The detection efficiency for Z=1 particles is the most important character for the PSD.
From the trajectory of the incident cosmic ray particle, we can get the module number it hits, and we regard this particle as detected if the signals from both ends of this module are larger than the threshold.
Fig.~\ref{fig:fig18} shows the result we got for all the detector modules. We can see that all the modules have a detector efficiency better than 99\% except few ones at the edge of the PSD, and the relative small efficiencies for those special ones are many due to their poor statistics.

\begin{figure}
 \centering
 \includegraphics[width=70mm]{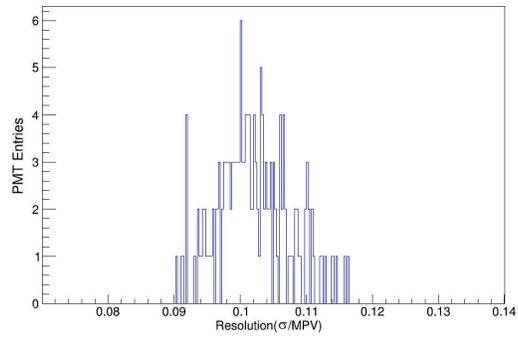}
 \caption{The charge resolution of the PSD detector modules obtained from the cosmic ray test.}
\label{fig:fig19}
\end{figure}

Charge resolution is also very important for the PSD. With the cosmic ray test, we also get this for Z=1 particles. For Z=1 particles, the charge resolution is equivalent to the energy resolution, and can be expressed as the ratio between the width ($\sigma$) and the MPV of the measured energy spectrum. The results is shown in Fig.~\ref{fig:fig19} and the charge resolution is better than 12\% for all the PMTs.

\begin{figure}
 \centering
 \includegraphics[width=70mm]{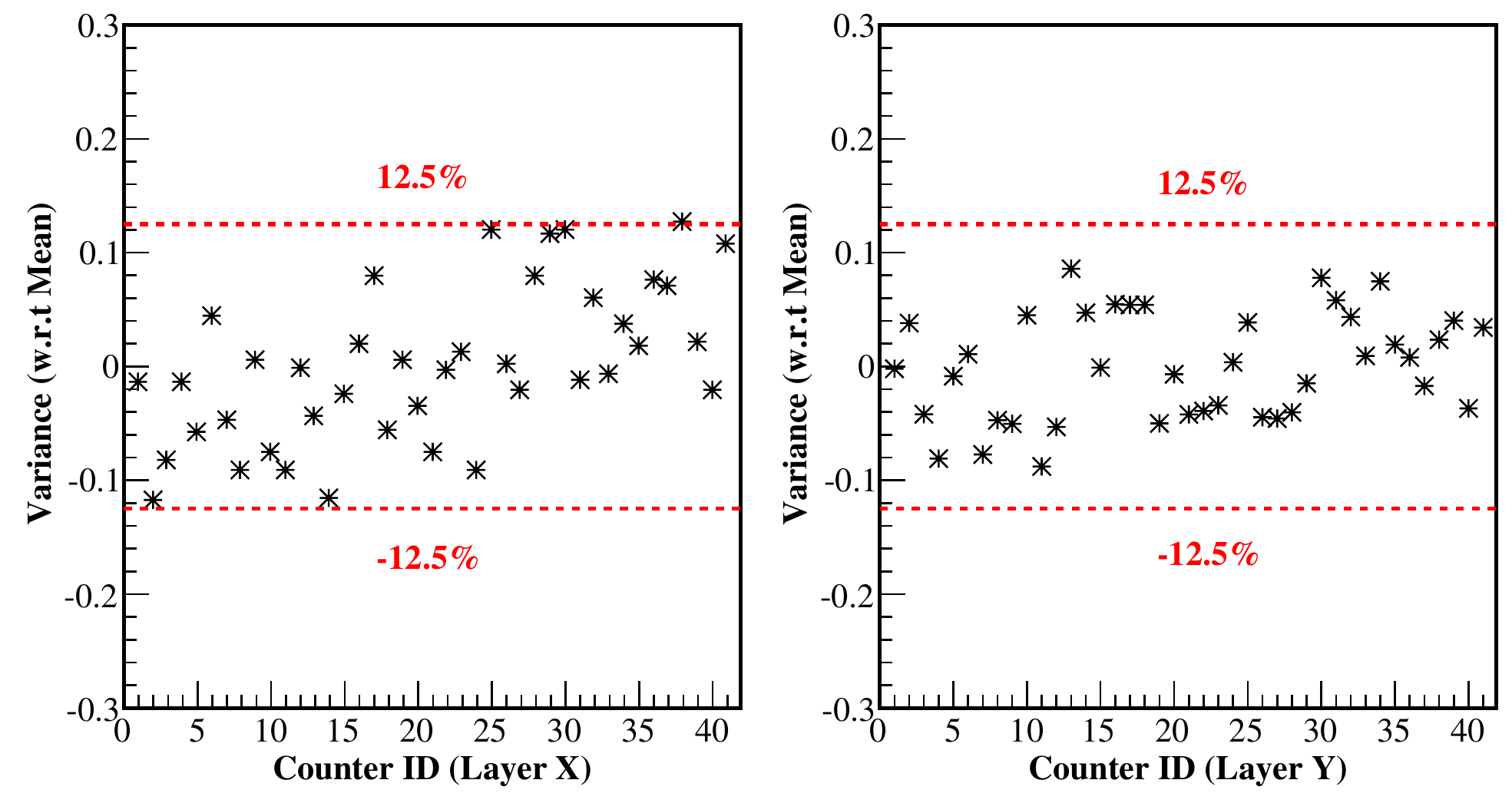}
\caption{The uniformity of the PSD detector modules obtained from the cosmic ray test.}
\label{fig:fig20}
\end{figure}

To check the consistency of the detector modules, we consider the average of the signals from both ends of a module, which will have a much weaker dependent on the incident position according to Fig.~\ref{fig:fig9}, to suppress the influence of different technical attenuation length. Fig.~\ref{fig:fig20} shows the ratio of difference to the average for all the modules, and only the particles incident on the center part of the modules are considered in the analyze. We can see that the largest difference between two detector modules is about 25 \%, and can satisfied the requirement for PSD.

\begin{figure}[!ht]
	\centering
	\includegraphics[width=40mm]{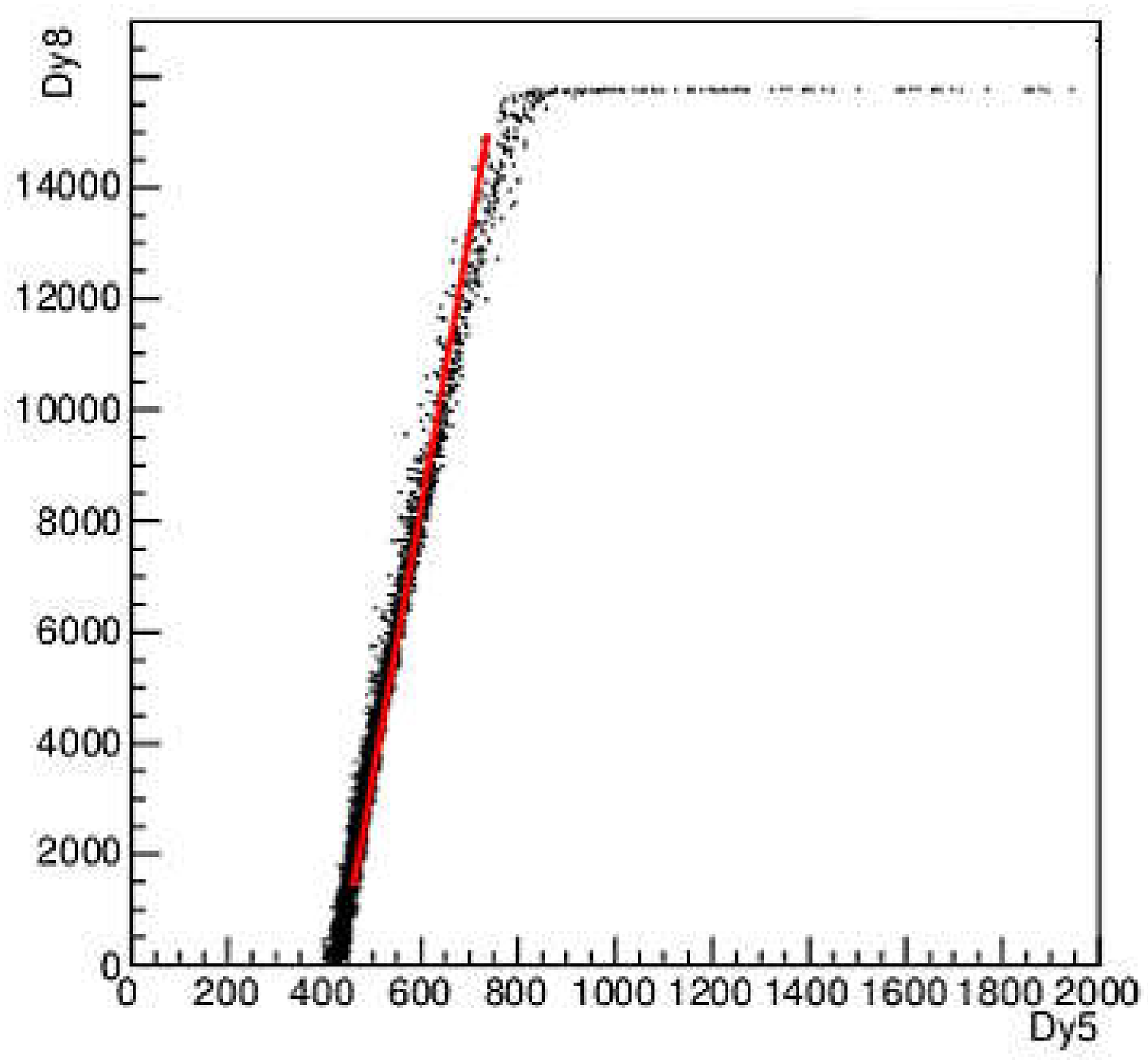}
	\includegraphics[width=40mm]{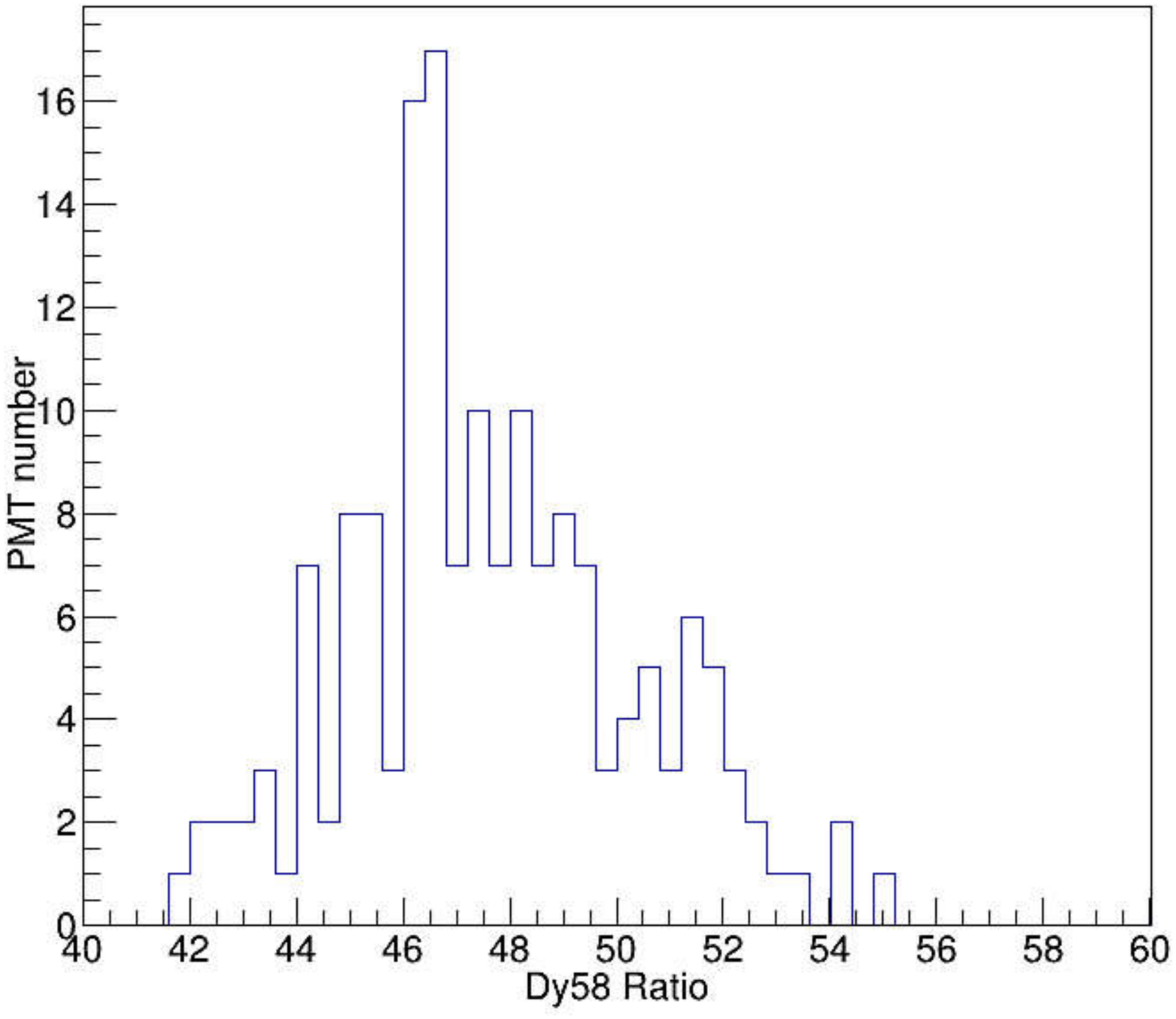}
	\caption{(a)A typical Dy8 vs Dy5 scatter plot for one PMT obtained from the cosmic ray test.(b)The Dy5/Dy8 gain ratio distribution of the PSD modules.}
     \label{fig:fig21}
\end{figure}

Because the limited technical attenuation length of the plastic scintillator bar, even with the cosmic ray, a broad large part of the Dy8 dynamic range can be covered due to the incident position varied. Fig.~\ref{fig:fig21}a shows the Dy5 vs Dy8 scatter plot for one PMT, and the gain ratio of Dy5 and Dy8 can be extracted from it. The obtained gain ratio distribution for all the PMTs is shown in Fig.~\ref{fig:fig21}b, and the values are within the range of 48 $\pm$ 7 for all the PMTs.

From the cosmic test, we know that the MPV for MIP signal is about 400 ADC channels, and the sigma of the noise is only about 4 channels, So it's not difficult to measure smaller signals down to 0.1 MIP for the detector modules. For larger signals, since we used a 14-bit ADC for analog digital conversion in the FEE, we can roughly estimate the dynamic range for each PMT by using:
\begin{equation}
Range_{max}=\frac{ADC\ Range_{max}\ (chan)}{Peak\ value\ of\ MIP\ (chan)}\times k_{Dy_{58}}
\end{equation}
where $k_{Dy_{58}}$ is the gain ratio of Dy5 and Dy8. Results shows that this value is larger than 1500 MIPs for any PMT we used, and this is fulfilled the requirement.

An accelerator beam test was carried on the EQM model of the PSD to verify our estimation in April 2015. The secondary isotopes with A/Z=2, which are produced with a 40 AGeV/c $^{40}{Ar}$ beam offered by CERN SPS, were used during this test. Preliminary results shows that when bombarding on the center of the modules, the signals for Ar is about 2500 channels in Dy5, and it will be 4000 for Fe by a simple extrapolation with the Birks-Chou law. The charge identification ability was also checked during the test, and it increases from 0.21 for He to 0.48 for Ar. These results validates that the PSD can cover and identify ion species up to Iron (Z = 26) in the mission, and more details will be presented in another paper.

As a space-borne apparatus, survival from the harsh environment of the mission, especially the launch phase, is the basic requirement for the PSD, and series of mechanics and thermotics tests were arranged to qualify this.

There are two standards for these tests, one is the acceptance level, which is already amplified by a factor to the real case to assure a safe margin of operation, and another is the authentication level, which is amplified again to prove the reliability of the design. To avoid the potential damage may caused, the FM of PSD only need to pass the acceptance level, and the EQM would go through those more severer tests also.

\begin{table}[!hbp]
	\centering
	\newcommand{\tabincell}[2]{\begin{tabular}{@{}#1@{}}#2\end{tabular}}	
	\caption{Conditions for mechanics tests} \label{tab:mechanics}
    \tabcolsep=8pt
	\begin{tabular}{c  c  c  c}
		\hline
		\multicolumn{2}{c}{\multirow{3}{*}{Projects}} & \multicolumn{2}{c}{Conditions} \\
		\cline{3-4}
		 & & \tabincell{c}{Acceptance \\ Level} &  \tabincell{c}{Authentication \\ Level} \\
		\hline
		\multirow{4}{*}{ \tabincell{c}{Sinu- \\ soidal}} & 5-8 Hz & 1.24 mm & 1.86 mm \\
		 & 8-100 Hz & 6 g max. & 9 g max. \\
		\cline{2-4}
		 &  \tabincell{c}{Sweep \\ Speed} & 4oct/min & 2oct/min \\ \hline
		\multirow{5}{*}{Random} & 20-100Hz &\multicolumn{2}{c}{+3 dB/oct.} \\
		 & 100-600Hz & 0.05$g^{2}$/Hz & 0.1$g^{2}$/Hz \\
		 & 600-2000Hz & \multicolumn{2}{c}{-9 dB/oct.} \\
		\cline{2-4}
		 & Grms & 6.41 & 9.07 \\
		\cline{2-4}
		 & Duration & 1 min & 2 min \\ \hline
		\multirow{3}{*}{Impulse} & 100-400 Hz & \multicolumn{2}{c}{+8 dB/oct.} \\
		 & 400-4000 Hz & \multicolumn{2}{c}{500 g} \\
		\cline{2-4}
		 & Time & 1  & 2 \\ \hline
	\end{tabular}
\end{table}

\begin{figure}
 \centering
 \includegraphics[width=70mm]{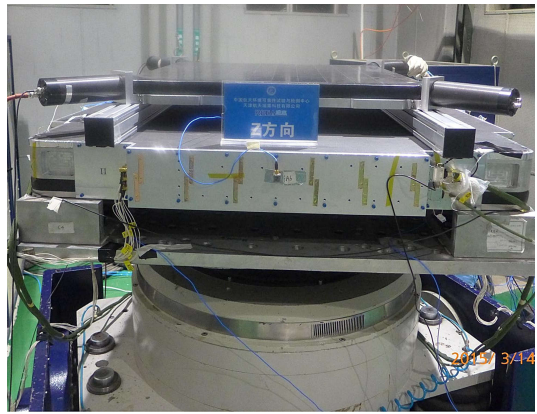}
\caption{Performance recheck with the cosmic rays for the PSD after a random vibration test run.}
\label{fig:fig22}
\end{figure}

The mechanics tests included the sinusoidal vibration, the random vibration and the impulse test, both for X/Y/Z 3 directions separately, and the conditions for these tests are giving in Table \ref{tab:mechanics}. No damage and deformation were detected after these tests, and the first resonance frequency of PSD was measured to be 166 Hz and 175 Hz for X/Y and Z direction respectively, well above the 128 Hz from the finite element analyzing.
We performed a performance check with cosmic ray before and after each test run, as shown in Fig.\ref{fig:fig22}, and no discrepancies on performance were observed.

\begin{figure}
 \centering
 \includegraphics[width=70mm]{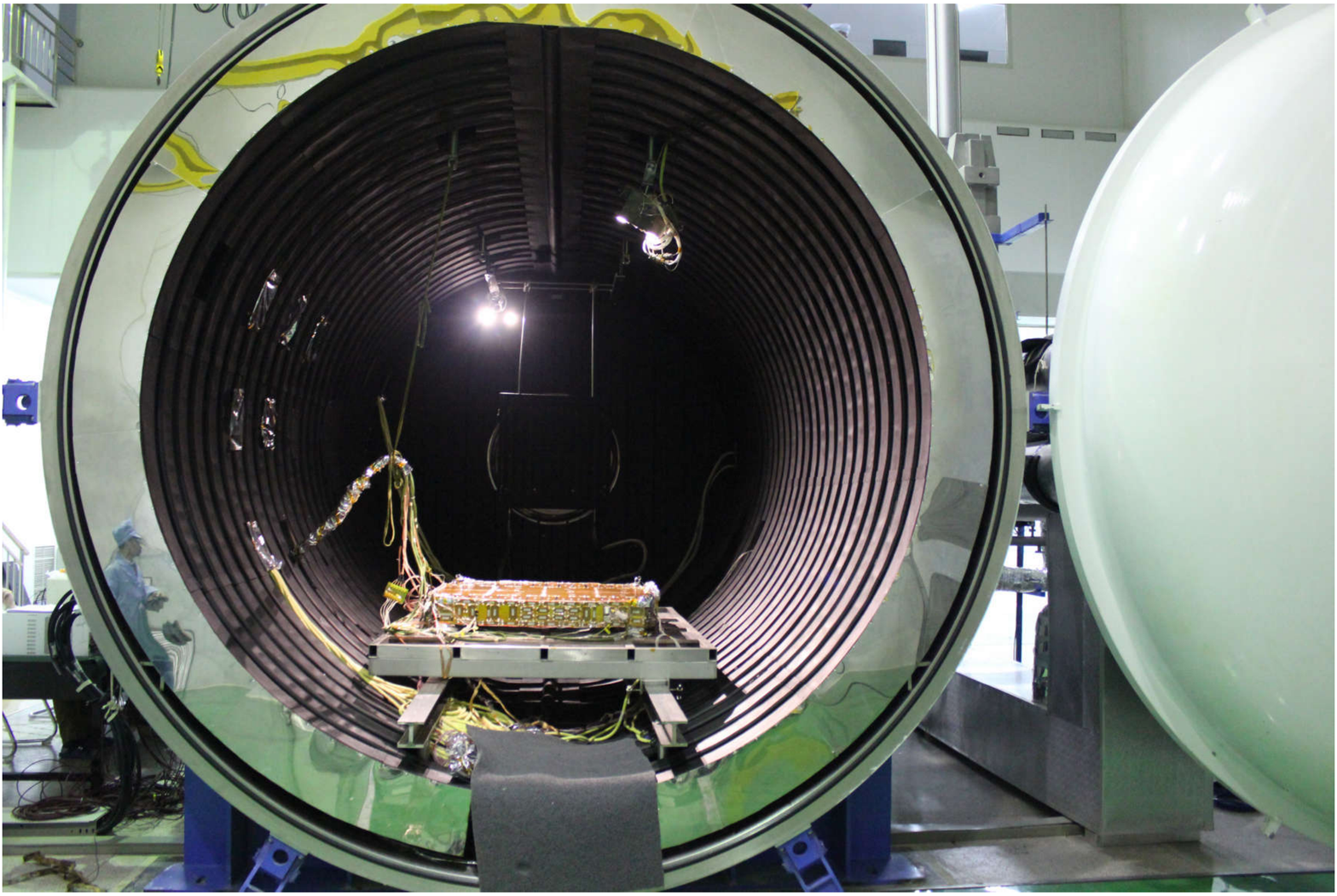}
\caption{The PSD in the vacuum chamber for the thermal vacuum test.}
\label{fig:fig23}
\end{figure}

Two thermotics qualification tests, the thermal cycle and the thermal vacuum test, have been performed. The PSD would underwent 25.5 cycles in a climatic chamber for the thermal cycle test. In each cycle, the temperature varied between -15 $^{\circ}$C and +35 $^{\circ}$C with a rate of 2 $^{\circ}$C/min and holden at every extreme value for about 5 hours. For the thermal vacuum test, the PSD was put into a large vacuum chamber, which had a vacuum better than 6.65 $\times$ $10^{-3}$ Pa as shown in Fig.\ref{fig:fig23}, and performed 6.5 cycles between -15 $^{\circ}$C and +35$^{\circ}$C. The temperature change rate was 1$^{\circ}$C/min and the PSD stayed at each extreme temperature for about 4 hours during the test. All cycles were passed successfully. The top cap was removed after each test for a inspection inside, and no damage and discrepancies were observed. During the tests, we also performed a cosmic ray test everytime the PSD kept in a extreme temperature in the thermal cycle test or throughout the whole process in the thermal vacuum test, and found no abnormal results. In a word, for all the qualification items, the PSD withstood the tests without damages or worsening of the performance.

Except these, a series tests for ElectroMagnetic Compatibility (EMC) had also been performed according to the national standard, and the results shows that the PSD has a strong capacity of resisting disturbance and will not affect other equipments in the satellite.

\section{Conclusions}

The plastic scintillator detector for the DAMPE has been successfully designed, built and tested. All the performance are satisfied with the requirements and it passed all the qualification tests successfully.
The flight model of PSD have been shipped and integrated into the DAMPE satellite in later April, 2015, and it shows a good reliability and stability in all the tests of the whole satellite system, which lasted for about half a year. The satellite has been successfully launched on December 17, 2015 for an at least 3-years mission, and the PSD will play a important role in it especially for the particle charge measurement and the photons/electrons separation.

%%%%%%%%%%%%%%%% Acknowledgement %%%%%%%%%%%%%%%%%%%%%%%
\section*{Acknowledgement}

This work is supported by the Strategic Priority Research Program on Space Science of the Chinese Academy of Science, Grant No. XDA04040202- 3 and the Youth Innovation Promotion Association, CAS. The authors wish to thank all the people from the DAMPE collaboration who helped to make this work possible.

%%%%%%%%%%%%%%%%   Bibliography  %%%%%%%%%%%%%%%%%%%%%%%
%% bibliography style
%%\section*{References}
%%\label{sec:reference}
%%\bibliographystyle{elsarticle-num}
%% From BibTex file
%%\bibliography{mybib}

%%\end{document}

%%%%%%%%%%%%%%%%   Bibliography  %%%%%%%%%%%%%%%%%%%%%%%
%% bibliography style
\section*{References}
%%\label{sec:reference}
%%\bibliographystyle{elsarticle-num}
%% From BibTex file
%%\bibliography{mybib}

\begin{thebibliography}{99}

\bibitem{Rubin} Rubin V C，Ford W K, {\it Aerospace Power Journal} {\bf 159}(1970) 379.
\bibitem{Clowe}	Clowe D, Bradač, Gonzalez A H，et al., {\it Aerospace Power Journal} {\bf 648} (2006) 109.
\bibitem{Klasen} M.Klasen, M.Pohl, G.Sig, {\it Progress in Part. and Nucl. Physics}{\bf 85} (2015) 1.
%%[10]	http://map.gsfc.nasa.gov/
%%[11]	http://www.sdss.org/
%%[12]	J P Ostriker, P Steinhardt Science 2003;300:1909-1913

\bibitem{chang} Chang Jin, {\it Chin. J. Space Sci.} {\bf 34} (2014) 550.
\bibitem{P.Azzarello} P.Azzarello et al., {\it Nuclear Inststrument and Methods in Physics Research A} {\bf 831} (2016) 378.
\bibitem{BGO_ECAL} zhiyong Zhang et al., {\it Nuclear Inststrument and Methods in Physics Research A} {\bf 836} (2016) 98.
\bibitem{scintillator} http://www.eljentechnology.com/
\bibitem{ACD_GLAST} A.A.Moiseev et al., {\it ASTROPART PHYS} {\bf 27} (2007) 339.
\bibitem{AMS_TOF} V.Bindi et al., {\it Nuclear Inststrument and Methods in Physics Research A} {\bf 623} (2010) 968.
\bibitem{r4443} http://www.hamamatsu.com
\bibitem{bethe} H. Bethe, J. Ashkin,{\it Experimental Nuclear Physics, ed. E. Segré, J. Wiley, New York}, {\bf 253} (1953).
\bibitem{dwyer85} R. Dwyer, D. Z. Zhou, {\it Nuclear Inststrument and Methods in Physics Research A}, {\bf 242} (1985) 171.
\bibitem{yong} Yong Zhou et al., {\it Nuclear Inststrument and Methods in Physics Research A} {\bf 827} (2016) 79.
\bibitem{airborn}http://www.airborn.com
\bibitem{VA_chip} Integrated Detector Electronics AS (IDEAS), VA160 datasheet, ( http://www.ideas.no)
\bibitem{tyvek} http://www.dupont.com/
\bibitem{zhang2015} Zhang yongjie et al., {\it NUCLEAR TECHNIQUES} {\bf 8} (2015) 080403
\bibitem{taiuti} Taiuti et al., {\it Nuclear Inststrument and Methods in Physics Research A} {\bf 27} (1996) 429.
\bibitem{zhou2016} Yong Zhou et al., {\it NUCL SCI TECH} {\bf} (2016) 70
\bibitem{rtv} http://www.momentive.com

\end{thebibliography}

\end{document}